\documentclass[draft]{agujournal2019}
\usepackage{url} 
\usepackage[finalnew]{trackchanges} 
\usepackage{soul}
\usepackage{amsmath}
\journalname{JGR: Solid Earth}

\begin{document}

%
%

\title{Nutation Damping from Core-Mantle Boundary Topography}

%
%




\authors{J.~Rekier\affil{1,2}, S.~A.~Triana\affil{1,2}, A.~Barik\affil{3}, D.~Abdulah\affil{4}, W.~Kang\affil{4}}

\affiliation{1}{Earth and Life Institute, UCLouvain, 1348 Ottignies-Louvain-la-Neuve, Belgium}
\affiliation{2}{Royal Observatory of Belgium, 3 avenue circulaire, 1180 Brussels, Belgium}
\affiliation{3}{Johns Hopkins University, 3400 North Charles Street, Baltimore, MD 21218, USA}
\affiliation{4}{Earth, Atmospheric and Planetary Science, Massachusetts Institute of Technology, USA}





\correspondingauthor{J.~Rekier}{jeremy.rekier@uclouvain.be}



\begin{keypoints}
\item Damping of the Free Core Nutation cannot be fully explained by electromagnetic coupling given current knowledge of the deep Earth.
\item Internal waves excited by the tidal flow over core-mantle boundary topography provide an additional source of dissipation.
\item A topography of typical amplitude 5 km dominated by features of wavelength 1500 km can fully account for the observed damping. 
\end{keypoints}

%
%

%
%


\begin{abstract}
Periodic gravitational forcing by the Moon and Sun produces small oscillations in Earth's rotation known as nutations. Nutations are amplified by a resonance with a natural motion of the liquid core called the Free Core Nutation, whose amplitude is limited by friction-like processes at the core--mantle boundary. Previous studies have attributed this damping to the dissipation of electric currents induced in the lower conducting mantle, but, given current knowledge of the lower mantle, electromagnetic coupling appears insufficient to fully account for the observed lag. We show that additional dissipation arises from the interaction of the tidal flow inside the core with the topography of the core--mantle boundary, which excites internal waves that extract energy and momentum from the flow. Adapting a theory originally developed for tides over seafloor topography, we find that the observed damping can be fully accounted for by a topography of typical amplitude $\sim$5~km dominated by features of wavelength $\sim$1500~km. The dissipation is highest when the upper core is neutrally buoyant. Such amplitudes are larger than typical inferences from global seismic studies but are not ruled out, given regional seismic evidence for kilometer-scale features and the sparse constraints at these scales.
\end{abstract}

\section*{Plain Language Summary}
The Earth's rotation axis wobbles slightly in space, a motion called nutation that is driven by the gravitational pull of the Moon and the Sun. Because the Earth has a liquid iron core inside a rocky mantle, this wobble is not perfectly rigid: the core sloshes back and forth relative to the mantle, and a small amount of energy is dissipated at the boundary between them, deep inside the Earth. Measuring this energy loss gives us a rare window into conditions at the core–mantle boundary. The current best explanation attributes the dissipation to electric currents flowing through the rocks just above the core, but this interpretation is recognized as imperfect, as it relies on properties of the deep Earth that are difficult to justify. We propose that an additional source of dissipation comes from tiny waves stirred up inside the core when the sloshing flow encounters bumps and ridges on the underside of the mantle. By adapting a theory developed for ocean tides flowing over seafloor mountains, we estimate the size and shape of these bumps. Our results suggest the core–mantle boundary is rougher than many global studies have found, but consistent with regional ones.

%
%

%


%
%
%
%

\section{Introduction}

Dissipative processes contribute to dampen the resonance of the Earth's nutation with the rotational mode known as the Free Core Nutation (FCN), causing the observed phase-lag between the tidal forcing and the Earth's response \cite{DehantMathews2015,MathewsEtAl2002}. Most of this occurs at the Core-Mantle Boundary (CMB), though some damping also occurs at the Inner-Core Boundary (ICB). Overall, this damping is relatively weak, measured with a quality factor at approximately $Q_\mathrm{FCN}\approx 20,000\pm2500$ \cite{VondrakRon2017,NurulHudaEtAl2020,ZhuEtAl2021,ZhangShen2021a}. In physical units, this corresponds to a total dissipated power of $P\approx 8\times10^{6}\,\mathrm{W}$ at the CMB (see Sec.~\ref{sec:FCN}). Dividing by the CMB surface gives an average power flux of $\mathcal{P}\approx5\times10^{-8}~\mathrm{W/m^2}$. Though small, this value has proven challenging to explain physically. Current models attribute it to Ohmic dissipation of electric currents in the lowermost mantle, a phenomenon referred to as Electromagnetic (EM) coupling \cite{Buffett1992,BuffettEtAl2002,MathewsGuo2005}. This interpretation relies on a conductivity of the lowermost mantle comparable to that of the upper core, which are generally disfavored by mineral physics studies \cite{Tyburczy2015}. While higher values might be possible \cite{OhtaEtAl2012,OhtaEtAl2014,HoEtAl2024}, an alternative dissipative mechanism is desirable to bridge the gap between theory and observations. 

Three decades ago, following early seismic assessments of the non-hydrostatic shape of the CMB, \citeA{WuWahr1997} began examining the effect of CMB topography on the amplitude of the Earth nutation, and variations in the length of day (LOD). \citeA{PuicaEtAl2023}, and \citeA{DehantEtAl2025} recently improved on this study based on more recent topographic models and found the correction on the nutation amplitude to be negligible. They did not consider the correction to the dissipation, a limitation that the present study aims to address. \citeA{LeMouelEtAl2006} also investigated dissipation at the CMB driven by topography, considering grain scales of centimeters to tens of meters produced by geochemical diffusion and crystallization (see also \citeA{MandeaEtAl2015}). They showed that small-scale topography can contribute to the dissipation by enhancing the viscous drag between the core and the mantle. At these scales, the topography cannot be constrained by seismological observations. The model presented here relies on a different mechanism operating on larger scales closer to the domain of seismic constraints.

Our knowledge of the CMB topography is currently very limited. Direct seismic measurements have large uncertainties and conclusions from various studies are not always compatible with each other. \citeA{Koelemeijer2021} performed a comprehensive review of the literature highlighting discrepancies between reconstructions from body waves and normal modes. Most studies, however, agree on a typical amplitude of the order of a few kilometers on scales commensurable with the CMB circumference, and these inferences are compatible with simulated profiles of dynamic topography caused by mantle convection \cite{Heyn2020}. There is considerably more uncertainty regarding the smaller scales. Some seismic studies have inferred kilometer-high topography at scales as small as a few hundred to a thousand kilometers. Recently, \citeA{MonvilleEtAl2025} computed the effect of CMB topography on the variations in the Length of Day (LOD) on decadal timescales. The magnetic Lorentz force is expected to contribute significantly to the momentum balance at these timescales, but very little on the diurnal timescale considered in the present work \cite{Braginsky1999}\add{, where ohmic dissipation in the bulk of the core is likewise negligible (Triana et al., 2021) and the boundary layer contribution is the one accounted for by the electromagnetic coupling theory reviewed in Sec.~4}. These authors used a model of topography synthesizing the data from seismic body waves assembled by \citeA{Koelemeijer2021}, and mantle simulations. We use a model of topography characterized by a similar power spectrum in the present paper (see Sec.~\ref{sec:Topomodel} below).

\begin{figure} 
	\centering
	\includegraphics[width=0.8\textwidth]{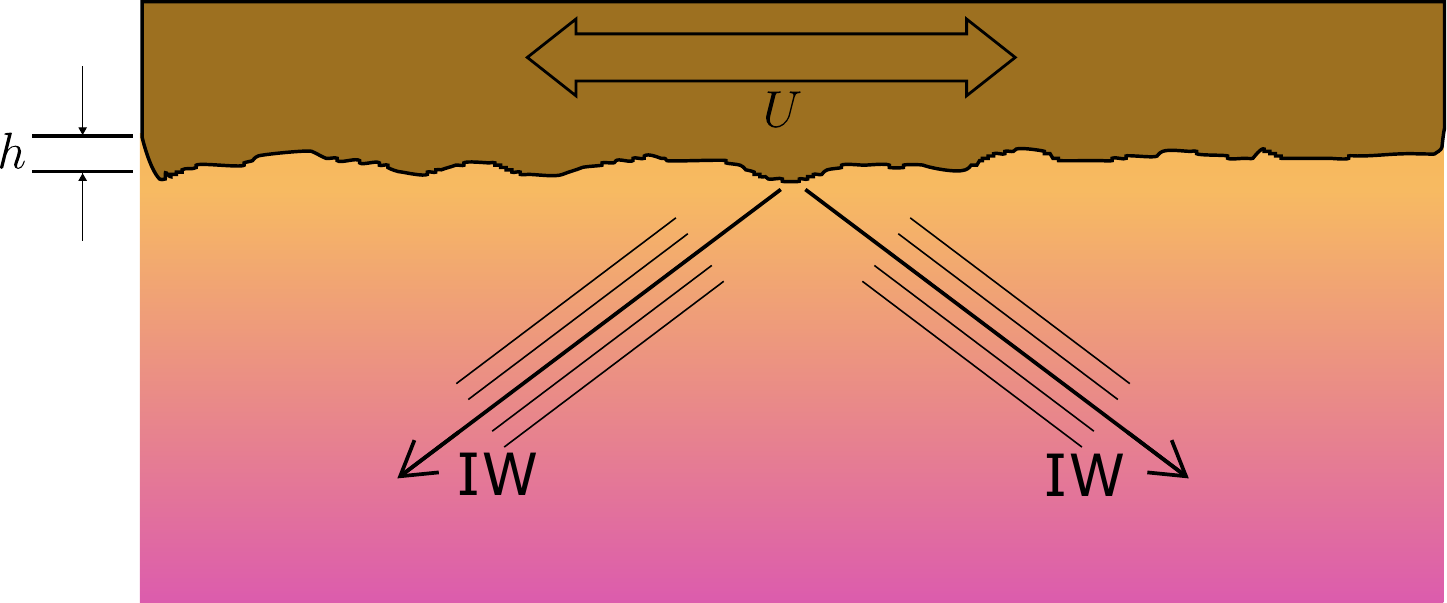}
	\caption{Sketch of the internal waves excited by the topographic features of the Core-Mantle Boundary. The nutation of the Earth's core produces a (quasi) diurnal tidal flow that interacts with the topography, exciting internal waves. These waves transport momentum which produces a periodic pressure drag on the surface and dissipates energy from the tidal flow. 
    }
	\label{fig:cmb_cartoon} 
\end{figure}

Our physical model is directly inspired by oceanography. In oceans, tidal flows interacting with seafloor topography excite internal waves which transport momentum upward, dissipating a fraction of the tidal energy into the ocean, and driving bottom mixing \cite{Bell1975,Bell1975a}. A similar process likely takes place at the Earth's CMB, where studies indicate the presence of a kilometer-high topography. In the Earth's core case, the tidal flow results from the differential rotation of the core relative to the mantle and has a (quasi) diurnal period. The oscillation of that flow over the topography excites internal waves that carry momentum away from the boundary and into the core, causing a pressure difference between the leading and the trailing sides of the topographic feature at the origin of the excitation. This results in a net pressure drag, aligned with the tidal flow velocity and proportional to its amplitude. At the global scale, this drag produces a torque on the mantle that dissipates a portion of the energy of the core differential rotation into the internal waves (Figure~\ref{fig:cmb_cartoon}). 

The analogy between Earth's oceans and its liquid core is not new. Braginsky employed it perhaps most explicitly when he introduced the concept of the \textit{Stratified Ocean of the Core} (SOC) \cite{Braginsky1993,Braginsky1999}. The flow in this ``hidden core ocean'' follows dynamical equations similar to those governing the flow at the seafloor. The existence of a stably stratified region at the top of the Earth's core has been an object of debate for more than fifty years \cite{Higgins1971,Whaler1980}. There have been several estimates of the properties of this layer. Studies of geomagnetic secular variations suggest a thickness $d\approx 100-140 ~\mathrm{km}$ \cite{gubbins_geomagnetic_2007,buffett_geomagnetic_2014}, and a buoyancy frequency--a.k.a. \textit{Brunt-V\"ais\"al\"a} frequency, $N$, of the order $N/\Omega\approx1.5-2$, where $\Omega=7.27\times10^{-5}\,\mathrm{s}^{-1}$ is the Earth's spin rate \cite{Braginsky1993,buffett_geomagnetic_2014}. Seismic evidence suggests $d\approx 100-450 \,\mathrm{km}$ \cite{Tanaka1993,Helffrich2013,Kaneshima2018,Tang2015}. Models of core thermodynamics and chemical barodiffusion \cite{lister_stratification_1998,Gubbins2013} point to a depth of the order $d\approx 100~\mathrm{km}$ with the frequency of $N/\Omega\approx1.4-20$. A recent numerical model of core convection taking into account thermal heterogeneity at the CMB finds regional stratification on top of the Earth's core with thickness of several hundred kilometers and $N/\Omega \approx 0.02-2$ \cite{Mound2019}. Numerical models of the Earth's geodynamo lean towards weaker stratification. Some models have suggested a stable layer with $d\approx100~\mathrm{km}$ with a frequency of the order of the spin rate, $N\approx\Omega$ \cite{Christensen2018}. More recent geodynamo simulations favor a very thin, weakly stratified layer, if one is present at all \cite{Gastine2020,Aubert2025}. In what follows, we treat the degree of the density stratification as an adjustable parameter.

In Sec.~\ref{sec:FCN}, we lay out the basic details concerning the FCN and its dissipation and compute the associated power flux at the CMB. In Sec.~\ref{sec:EMcoupling} we then briefly review the EM coupling mechanism based on recent estimates of the magnetic field intensity and mantle conductivity. We describe our mathematical formalism in Sec.~\ref{sec:Topocoupling}, and our topographic model in Sec.~\ref{sec:Topomodel}. In Sec.~\ref{sec:Results}, we infer constraints on the topographic amplitude and density stratification at the CMB. 

\section{The Free Core Nutation and its dissipation}
\label{sec:FCN}

The Free Core Nutation of the Earth is a free rotational mode with frequency \cite{DehantMathews2015,Rekier2022a}:
\begin{equation}
    \omega_\mathrm{fcn}=-\Omega-\left(1+\frac{A_f}{A_m}\right)\left(e_f-\beta+K_\mathrm{cmb}+\frac{A_s}{A_f}K_\mathrm{icb}\right)\Omega,
    \label{eq:omegaFCN}
\end{equation}
as measured in the terrestrial frame of reference. The symbols $A_s$, $A_f$, and $A_m$, represent the equatorial moments of inertia of the solid inner core, fluid outer core, and mantle, respectively. The FCN frequency is retrograde and quasi-diurnal, $\omega_\mathrm{fcn}\approx-\Omega$, with a small correction given by the combination of terms enclosed in the second set of parentheses. Among these terms $e_f$ parametrizes the flattening of the CMB, and $\beta$ is a compliance parameter characterizing the elastic deformation properties of the mantle. The two remaining parameters, $K_\mathrm{cmb}$ and $K_\mathrm{icb}$ are complex valued. They parametrize the torque and the associated dissipation at the CMB, and at the ICB.

The kinetic energy of the FCN is due mostly to the uniform vorticity flow inside the core. A very good approximation of the rate of energy dissipation is then given by \cite{Buffett2010a}:
\begin{equation}
    \frac{dE}{dt}=A_f|\boldsymbol{\omega}_f|^2\mathrm{Im}(\omega_\mathrm{fcn}),
    \label{eq:dE/dt}
\end{equation}
where $\boldsymbol{\omega}_f$ is the angular velocity of the core relative to the mantle. The theory of nutation gives $|\boldsymbol{\omega}_f|\approx2.4\times10^{-11}\,\mathrm{s}^{-1}$ \cite{MathewsEtAl2002}. 
By inserting Equation~\eqref{eq:omegaFCN} into Equation~\eqref{eq:dE/dt}, we obtain the sum of a term proportional to $K_\mathrm{cmb}$, and one proportional to $K_\mathrm{icb}$. The first of these corresponds to the dissipated power at the CMB:
\begin{equation}
    P_\mathrm{cmb}=-A_f\left(1+\frac{A_f}{A_m}\right)\Omega~|\boldsymbol{\omega}_f|^2\mathrm{Im}(K_\mathrm{cmb}).
\end{equation}
Injecting $A_f=9.0583\times10^{36}\,\mathrm{kg\,m^2}$, and $A_m=7.0996\times10^{37}\,\mathrm{kg\,m^2}$, gives $P_\mathrm{cmb}\approx8\times10^{6}\,\mathrm{W}$, based on the value of $\mathrm{Im}(K_\mathrm{cmb})=-1.87\times10^{-5}$ inferred from observations \cite{RekierEtAl2021}. Dividing by the CMB surface places the average power flux at $\approx5\times10^{-8}~\mathrm{W/m^2}$.

\section{Review of Electromagnetic Coupling}
\label{sec:EMcoupling}
\begin{figure} 
	\centering
	\includegraphics[width=0.8\textwidth]{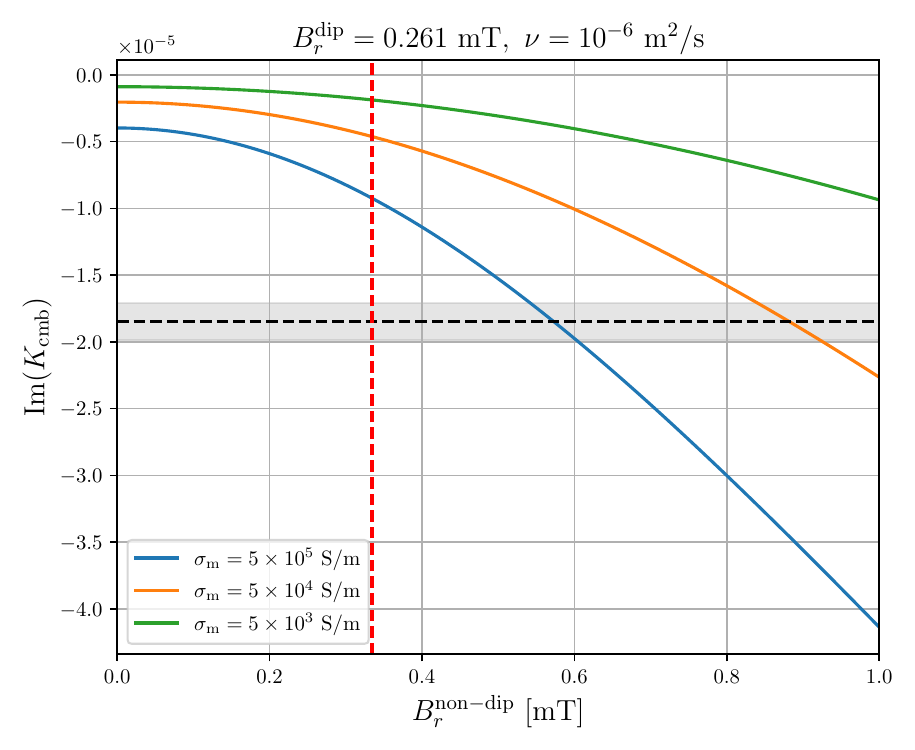}
	\caption{Imaginary part of the coupling constant in the EM coupling scenario. The colors correspond to 3 different values of the mantle conductivity. A rms value of $0.261\,\mathrm{mT}$ is assumed for the dipolar magnetic field. The vertical red line corresponds to a rms value of $0.334\,\mathrm{mT}$ for the non-dipolar part of the magnetic field inferred from the magnetic data. 
        The dashed line corresponds to the value of $\mathrm{Im}(K_\mathrm{cmb})$ inferred from the nutation data with uncertainty represented by the shaded area.
        }
	\label{fig:Kcmb} 
\end{figure}

We can compute the electromagnetic dissipation at the CMB using the theory of Mathews and Guo which also takes into account the much smaller contribution of viscous dissipation \cite{MathewsGuo2005}. Setting the kinematic viscosity of the core to its estimated value for liquid iron, $\nu=10^{-6}\,\mathrm{m^2s^{-1}}$, the result depends on the conductivity of the core, which is usually taken as $\sigma_\mathrm{f}=5\times10^{5}~\mathrm{S}\mathrm{m}^{-1}$ \cite{Olson2015}, and of the lower mantle, $\sigma_\mathrm{m}$. It also depends on the radial magnetic field at the CMB, $B_r$, which is parametrized as the sum of a dominant dipolar component, $B_r^\mathrm{dip}$, which can be easily inferred from surface observation, and a non-dipolar component, $B_r^\mathrm{non-dip}$ which parametrizes the collective contribution of smaller scale components \cite{BuffettEtAl2002,KootEtAl2010}. Figure~\ref{fig:Kcmb} gives a measure of the dissipation as a function of the radial magnetic field for three values of the mantle conductivity, with the core conductivity treated as an upper bound. We have used the \texttt{planetMagFields} software to obtain the dipolar and non-dipolar components of the magnetic field at the CMB from the 14th generation IGRF data \cite{BegganEtAl2026,BarikAngappan2024}.

\add{
The IGRF-14 main field model is truncated at spherical harmonic degree $\ell=13$ (Beggan et al., 2026), corresponding to a horizontal wavelength of about $1600$\,km at the CMB. The value of $B_r^\mathrm{non-dip}$ obtained from it therefore excludes any contribution from smaller scales, which are certainly present but cannot be recovered from surface observations. The geomagnetic spatial spectrum is close to white at the CMB over the observed range of degrees, and extrapolating it beyond the truncation gives $0.52\,\mathrm{mT}$ for a spectrum remaining white to $\ell=30$, or $0.68\,\mathrm{mT}$ to $\ell=50$, against the $0.334\,\mathrm{mT}$ resolved by IGRF. Numerical dynamos now reproduce the observed core surface field in detail (Aubert, 2026), and a determination of the CMB field spectrum beyond the observable degrees from such simulations would usefully constrain this extrapolation. The horizontal axis of Figure~2 spans the plausible range.

The figure shows that the two parameters trade off against one another. Accounting for the observed dissipation requires $B_r^\mathrm{non-dip}\approx0.58\,\mathrm{mT}$ if the lower mantle conductivity equals that of the core, but $\approx0.90\,\mathrm{mT}$ if it is ten times smaller, corresponding to white spectra extending to $\ell\approx37$ and $\ell\approx88$ respectively. Unresolved small-scale field thus relaxes the requirement on the mantle conductivity without removing it: at $\sigma_\mathrm{m}=5\times10^{3}~\mathrm{S}\mathrm{m}^{-1}$ the field would have to remain white beyond $\ell\approx200$, \textit{i.e.} about $100$\,km at the CMB. Explaining the nutation dissipation by electromagnetic coupling alone therefore requires a lower mantle conducting nearly as well as liquid iron, whatever the amplitude of the field at unresolved scales.

Whether such conductivities are plausible remains open. They are generally disfavoured by mineral physics studies (Tyburczy \& Du Frane, 2015), but independent geodetic evidence points the other way: the reflection of torsional waves and the amplitude of decadal length-of-day variations both favour a lower mantle conductance in excess of $10^{9}\,\mathrm{S}$ (Gillet et al., 2015; Aubert, 2026), which would require material of near-core conductivity distributed over a layer some tens of kilometres thick. We note that the two observables probe different depths, the electromagnetic skin depth being of order $200$\,m at the diurnal periods relevant here against tens of kilometres at decadal periods, so that a conductance inferred from length-of-day variations does not directly translate into the conductivity sampled by the nutations. Electromagnetic coupling therefore cannot be excluded, but accounting for the observed dissipation by that mechanism alone requires a combination of assumptions each at the limit of what is currently supported. This motivates the search for an additional dissipative mechanism, which we present in the following sections.}
  
\section{Topographic coupling}
\label{sec:Topocoupling}

We follow \cite{Bell1975a} and solve the equation of fluid dynamics linearized about a tidal base flow state. Choosing a local cartesian set of coordinates with the $z$-axis aligned with the Earth's radius, and the $x$ and $y$ axes pointing along the zonal and meridional directions, respectively, the dynamics is governed by the following equations \cite{Vallis-2017:atmospheric}:
\begin{subequations}
\begin{align}
\frac{\mathrm{D} u}{\mathrm{D} t}+\tilde{f} w-f v&=-\frac{1}{\rho} \frac{\partial p}{\partial x} \label{eq:momx}\\
\frac{\mathrm{D} v}{\mathrm{D} t}+f u&=-\frac{1}{\rho} \frac{\partial p}{\partial y} \label{eq:momy}\\
\frac{\mathrm{D} w}{\mathrm{D} t}-\tilde{f} u&=-\frac{1}{\rho} \frac{\partial p}{\partial z}+b\label{eq:momz}\\
\frac{\mathrm{D} b}{\mathrm{D} t}&=-N^2w\label{eq:buoyancy}\\
\frac{\partial u}{\partial x}+\frac{\partial v}{\partial y}+\frac{\partial w}{\partial z}&=0\label{eq:incompressibility}
\end{align}
\end{subequations}
where $f=2\Omega\sin{\vartheta}$, and $\tilde{f}=2\Omega\cos{\vartheta}$, with $\vartheta$ denoting latitude, and $\Omega$ the Earth's spin rate. 
\add{
The variable $b$ denotes the buoyancy, and $N$ the Brunt-V\"ais\"al\"a frequency introduced in Section 1. The model is purely hydrodynamic; the magnetic field does not enter, and the coupling computed here is independent of the electromagnetic mechanism reviewed in Section 3.
}
To linear order in the velocity, the material derivative reduces to:
\begin{equation}
    \frac{\mathrm{D}}{\mathrm{D} t}\equiv\left(\frac{\partial}{\partial t}+\mathbf{U}\cdot\mathbf{\nabla}\right),
\end{equation}
where $\mathbf{U}$ is the tidal base flow. Bell examined the one-dimensional harmonic forcing. We are interested in the tidal flow produced by a single retrograde nutation component of angular frequency, $\omega_0$, which can be written as \cite{Buffett2010a,ShihEtAl2023}:
\begin{equation}
    \mathbf{U}=U_0\sin\vartheta\cos{(\omega_0t+\varphi)}~\hat{\mathbf{x}}-U_0\sin{(\omega_0t+\varphi)}~\hat{\mathbf{y}},
    \label{eq:baseflow}
\end{equation}
where $\hat{\mathbf{x}}$, and $\hat{\mathbf{y}}$, are unit vectors along the zonal and meridional directions, respectively. The amplitude of the main nutation component is $U_0=8.7\times10^{-5}\mathrm{m/s}$. \citeA{Bell1975a} gave the first solution to Eqs.~\eqref{eq:momx} to \eqref{eq:incompressibility} subjected to the boundary condition applying at a rough surface:
\begin{equation}
    w|_{z=0}=\mathbf{U}\cdot\mathbf{\nabla}h,
    \label{eq:bc}
\end{equation}
where $h(x,y)$ is the \textit{elevation function} parametrizing the topography. Bell's original study neglected the rotation. It was reintroduced later in his application to the Earth's ocean \cite{Bell1975}. The latter relied on the \textit{Traditional Approximation} (TA) which assumes $\tilde{f}=0$. A simplification that amounts  to neglecting the vertical component of the Coriolis force and can only be considered valid when buoyancy dominates (i.e. $N\gg\Omega$). For small $N$, the TA breaks down near the equator. Since our analysis requires dissipation across all latitudes, we have generalized the model beyond the TA. All the technical details of this extension can be found in the Supplementary Information (SI). 

After solving, Eqs.~\eqref{eq:momx} to \eqref{eq:incompressibility}, we find the pressure drag on the CMB patch of area, $A$, as (we use the common short-hand $d^2x$ to denote the surface element throughout):
\begin{equation}
    \boldsymbol{\mathcal{F}}=\frac{1}{A}\int_Ap|_{z=0}\mathbf{\nabla}h~d^2x.
    \label{eq:formdragdef}
\end{equation}
The power flux is then equal to the dot product of the pressure with the tidal flow, which upon using the Eq.~\eqref{eq:bc} writes:
\begin{equation}
    \mathcal{P}=\frac{1}{A}\int_Awp|_{z=0}~d^2x.
    \label{eq:powerfluxdef}
\end{equation}
After a significant amount of algebra (see~\ref{sec:derivation}), the power flux averaged over one tidal cycle can be expressed as an integral in Fourier space:
\begin{equation}
    \bar{\mathcal{P}}=\frac{\rho}{2\pi^2}\sum_{n=1}^{\infty}(n\omega_0)\int \frac{\mathcal{S}(k_x,k_y)}{k}J_n(\beta)^2\sqrt{(n^2\omega_0^2 - N^2)(f^2 - n^2\omega_0^2) + n^2\omega_0^2 f_s^2}~d^2k~.
    \label{eq:powerflux}
\end{equation}
where $\mathcal{S}(k_x,k_y)$ is the \textit{topographic power spectrum} (see Sec.~\ref{sec:Topomodel}), and where $f_s=\tilde{f}\sin\alpha$, with $\alpha$ the angle between the direction of propagation of the wave, and the azimuthal direction. $k_x$ and $k_y$ are the wavenumbers in the $x$ and $y$ directions, respectively ($d^2k$ denotes the surface element in Fourier space). Equation~\eqref{eq:powerflux} is the sum over the individual contributions of internal plane waves, whose wave vectors components in the $x$ and $y$ directions are $k_x$ and $k_y$, respectively, and whose frequencies are multiples of the exciting tidal frequency which in our case is diurnal, $\omega_0=-\Omega$. The argument of the Bessel function, $J_n$, is defined as:
\begin{equation}
    \beta=\frac{kU_0}{|\omega_0|}\sqrt{\sin^2\vartheta\cos^2\alpha+\sin^2\alpha}.
\end{equation}
Since $U_0/|\omega_0|\approx1\,\mathrm{m}$, the so-called \textit{quasi-static approximation}, which requires $\beta\ll1$, is well justified for topographic features wider than that length scale, which further simplifies the power flux formula (see Sec.~\ref{sec:QSapprox} \add{for details}).

\section{Topographic model}
\label{sec:Topomodel}

The topographic power spectrum, $\mathcal{S}$, characterizing the roughness of a surface of area $A$ is defined by:
\begin{equation}
    h_\mathrm{rms}^2=\frac{1}{A}\int h(x,y)^2~d^2x=\int \mathcal{S}(k_x,k_y)~d^2k,
\end{equation}
where $h_\mathrm{rms}$ denotes the root-mean-square amplitude of the topography. For simplicity, we treat the power spectrum as isotropic, and we adopt the following model which has been shown to adequately capture the behavior of geophysical \cite{GagnonEtAl2006,CandelaEtAl2012,NaveiraGarabatoEtAl2013} as well as synthetic surfaces \cite{PerssonEtAl2004,JacobsEtAl2017,RodriguezEtAl2025}:
\begin{equation}
    \mathcal{S}(k)=
        \frac{h_\mathrm{rms}^2\mathcal{H}}{k_0^2\pi}\left(1+\frac{k^2}{k_0^2}\right)^{-(\mathcal{H}+1)}
    \label{eq:powerspectrum}
\end{equation}
where $k=\sqrt{k_x^2+k_y^2}$, and $\mathcal{H}$ is the \textit{Hurst exponent} which characterizes the degree of self-similarity of the surface across length scales and is related to the fractal dimension $D_f = 3 - \mathcal{H}$ \cite{PerssonEtAl2004,RodriguezEtAl2025}. At wavenumbers $k\gg k_0$, the spectrum decays as $\mathcal{S}(k)\sim k^{-2(\mathcal{H}+1)}$, so that a larger $\mathcal{H}$ implies a steeper decay and therefore a smoother surface at small scales. Values for natural surfaces are typically $0.7\leq\mathcal{H}\leq0.9$\cite{CandelaEtAl2012}, although values closer to $\mathcal{H}\approx0.6$ are also possible \cite{GagnonEtAl2006}. The last parameter, $k_0$, is crucial. It sets the horizontal wavelength, $\lambda_0=2\pi/k_0$ above which the power spectrum flattens out. In surfaces generated for industrial applications, this wavelength corresponds to the largest topographic feature \cite{PerssonEtAl2004}. The wavelength of the CMB topography is naturally bounded by the circumference of the Earth's core, and thus: $k_0\geq1/R$, with $R=3485\,\mathrm{km}$ being the core radius. 

Strictly speaking, the power spectrum $\mathcal{S}(k)$ is defined in a planar geometry, which is appropriate for the small-scale topographic features for which it was originally developed. Observational constraints on CMB topography, however, are exclusively available at the largest scales, and typically expressed in terms of spherical harmonic coefficients $h_\ell^m$. We therefore need to relate the planar spectral model to the spherical harmonic representation. For an isotropic field on a sphere, the total variance decomposes as:
\begin{equation}
    h_\mathrm{rms}^2 = \sum_{\ell=0}^{\infty} (2\ell+1) C_\ell,
    \label{eq:hrms_SH}
\end{equation}
where $C_\ell = \langle |h_\ell^m|^2 \rangle$ is the angular power spectrum, and the factor $(2\ell+1)$ counts the number of independent modes at each degree. In the planar representation, the same variance is recovered by integrating the power spectrum over all wavenumbers:
\begin{equation}
    h_\mathrm{rms}^2 = \int_0^\infty 2\pi k\, \mathcal{S}(k)\, dk.
    \label{eq:hrms_k}
\end{equation}
Identifying degree $\ell$ with wavenumber $k_\ell = \ell/R$, adjacent degrees are separated by $\Delta k = 1/R$, and the variance contributed by a single degree in the planar representation is:
\begin{equation}
    2\pi k_\ell\, \mathcal{S}(k_\ell)\, \Delta k 
    = \frac{2\pi \ell}{R^2}\, \mathcal{S}(k_\ell).
\end{equation}
Equating this with $(2\ell+1)C_\ell \approx 2\ell\, C_\ell$ gives $C_\ell = \pi\, \mathcal{S}(k_\ell)/R^2$, and therefore:
\begin{equation}
    \sqrt{(2\ell+1)C_\ell} \approx \sqrt{2\pi k_\ell^2\, 
    \mathcal{S}(k_\ell)}.
    \label{eq:bridge}
\end{equation}
Both sides represent the topographic amplitude at a given scale: the left-hand side is what seismological studies report, which we relate to:
\begin{equation}
    h(k) = \sqrt{2\pi k^2\,\mathcal{S}(k)}.
    \label{eq:hk}
\end{equation}
The approximation $(2\ell+1)\approx 2\ell$ in eq.~\eqref{eq:bridge} introduces an error of at most $10\%$ for $\ell \geq 4$, which is small compared to the scatter in the available seismological estimates.

The seismological estimates of topographic amplitudes at low degrees, compiled by \citeA{Koelemeijer2021}, scatter over nearly an order of magnitude depending on the study and methodology. The more recent estimates, which we take as more reliable, suggest a peak-to-peak amplitude at $\ell=2$ of roughly $1$--$3\,\mathrm{km}$, growing to $4$--$8\,\mathrm{km}$ when summed to $\ell=4$. The fact that the amplitude at $\ell=4$ exceeds that at $\ell=2$ suggests that the peak of $h(k)$, occurring at $k=k_0$, lies at $k_0 \geq k_2$, but the data do not further constrain $k_0$ from above.

To calibrate the spectrum, we select an anchoring observation $(k_\star, h_\star)$ and invert eq.~\eqref{eq:hk} for $h_\mathrm{rms}$. In the particular case $k_0 = k_\star$, this gives:
\begin{equation}
    h_\mathrm{rms} = h_\star \sqrt{\frac{2^\mathcal{H}}{\mathcal{H}}},
    \label{eq:hrms_calibration}
\end{equation}
where the trailing factor varies little over the range $\mathcal{H}\in[0.6,\,1.5]$. As a reference case, we set $k_0 = k_\star = k_4 = 4/R$ and $h_\star \approx 2.5\,\mathrm{km}$, a mid-range estimate consistent with the recent literature, giving $h_\mathrm{rms} \approx 3.7\,\mathrm{km}$ for $\mathcal{H} = 0.8$.

\citeA{MonvilleEtAl2025} adopted a power spectrum similar to eq.~\eqref{eq:powerspectrum}, calibrated against a single study combining seismic data constrained by mantle flow simulations \cite{SoldatiEtAl2012}, which yields significantly smaller topographic amplitudes than our reference case. Fitting the spectrum given in their figure~1 to eq.~\eqref{eq:powerspectrum} gives $\mathcal{H}\approx1.3$, which exceeds the self-affine bound $\mathcal{H} \leq 1$ imposed by $D_f = 3 - \mathcal{H} \geq 2$ and corresponds to a surface smoother than any fractal. \change{This likely reflects the inherent damping of small-scale features in viscous mantle flow simulations. We note that the same figure shows that Earth's surface topography scales as $\mathcal{H}\approx0.93$, and that the corresponding simulated surface systematically underestimates the amplitude at small scales.}{We note that seismic tomography constrains the CMB only at the longest wavelengths, so that the small-scale end of such a spectrum reflects the flow model rather than the data.} Several of the regional studies reviewed by \citeA{Koelemeijer2021} infer kilometer-scale topography at wavelengths of a few hundred to a thousand kilometers, well below the scales resolved by such simulations. The spectrum of \citeA{MonvilleEtAl2025} is a valid possibility, but given the state of the observational constraints we choose not to privilege it over other options, and we treat $h_\mathrm{rms}$, $k_0$ and $\mathcal{H}$ as free parameters, anchored by but not restricted to the reference case above.

\section{Results}
\label{sec:Results}

\begin{figure}
    \centering
    \includegraphics[width=1\linewidth]{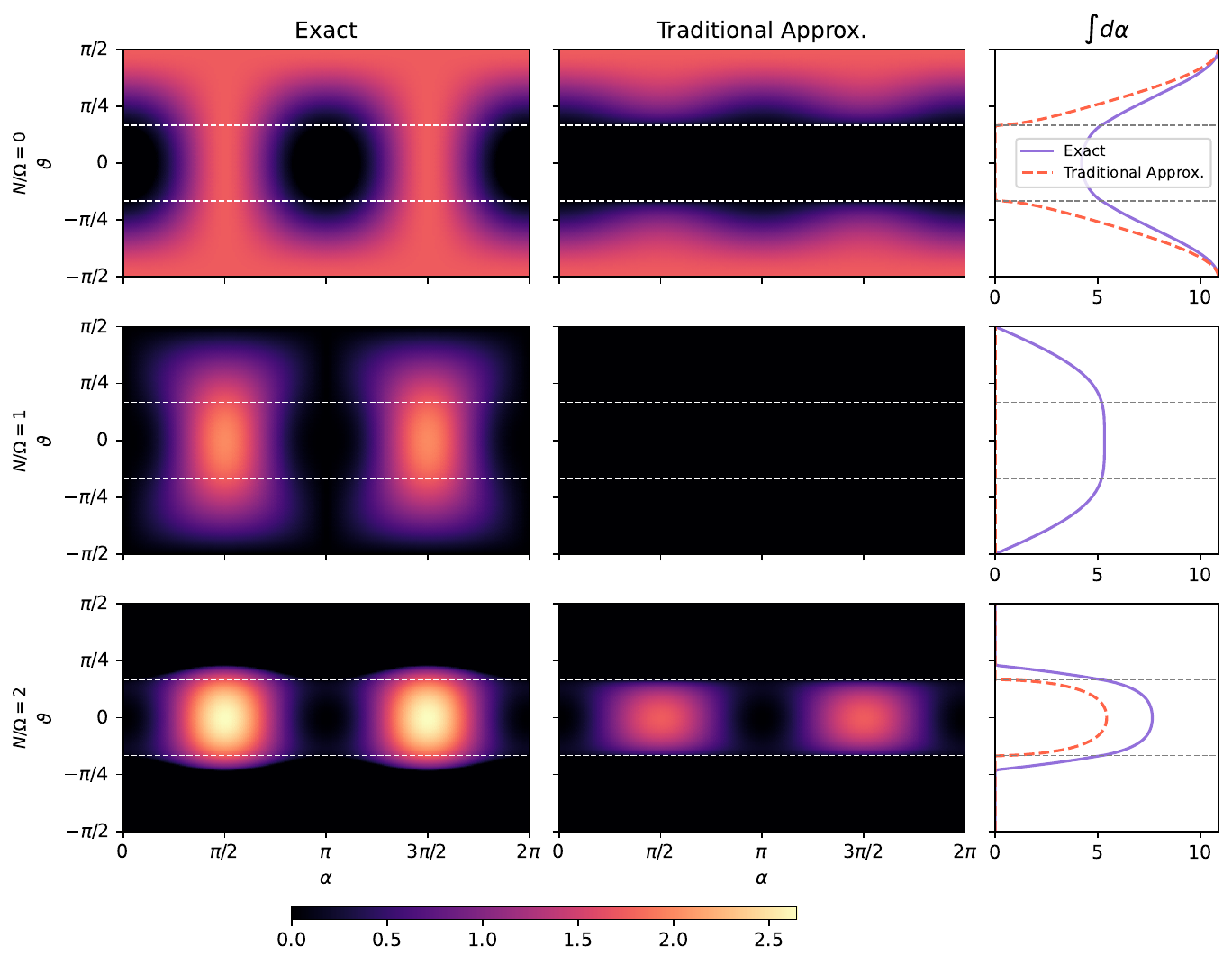}
    \caption{Left: integrand of eq.~\eqref{eq:powerfluxfinal} as a function of the $\vartheta$, and $\alpha$, for different values of $N/\Omega$. Center: same thing under the traditional approximation. Right: both cases after integrating over $\alpha$. The dashed lines indicate where $\sin\vartheta=1/2$.}
    \label{fig:flux_map}
\end{figure}

Upon inserting equation~\eqref{eq:powerspectrum} into the formula for the power flux derived in~\ref{sec:QSapprox}, and setting $\omega_0=-\Omega$, we obtain:
\begin{multline}
    \bar{\mathcal{P}}=\frac{\rho U_0^2\Omega~h_\mathrm{rms}^2k_0}{32\pi^{5/2}}\frac{\Gamma(\mathcal{H}-1/2)}{\Gamma(\mathcal{H})}\\
    \times\int_0^{2\pi}(\sin^2\vartheta\cos^2\alpha+\sin^2\alpha)\sqrt{\left(1 - \frac{N^2}{\Omega^2}\right)(4\sin^2\vartheta - 1) + 4 \cos^2\vartheta\sin^2\alpha}~d\alpha.
    \label{eq:powerfluxfinal}
\end{multline}
The leading factor contains the dependence of the power flux on the shape of the topography. The integral factor depends on the latitude. Equation~\eqref{eq:powerfluxfinal} is physically meaningful only when it evaluates to a real number, which requires the expression under the square root to be positive. The column on the left of Figure~\ref{fig:flux_map} shows the integrand of eq.~\eqref{eq:powerfluxfinal} as a function of the latitude, $\vartheta$, and $\alpha$, for different values of $N/\Omega$. The center column shows the same thing under the traditional approximation. In both columns, the white dashed line indicate where $\sin\vartheta=1/2$. The right column shows both cases after integrating over $\alpha$.
Two regimes emerge. When $N<\Omega$, the dissipation results from the excitation of \textit{inertial waves}, the dynamics of which is dominated by the effect of the Coriolis force. For $N>\Omega$, the dissipation results from the excitation of \textit{gravity waves}, the dynamics of which is dominated by the effect of buoyancy. Both types of waves become evanescent for $N=\Omega$ under the traditional approximation, as expected from the classic theory \cite{Vallis-2017:atmospheric}. 

\begin{figure}
    \centering
    \includegraphics[width=0.8\linewidth]{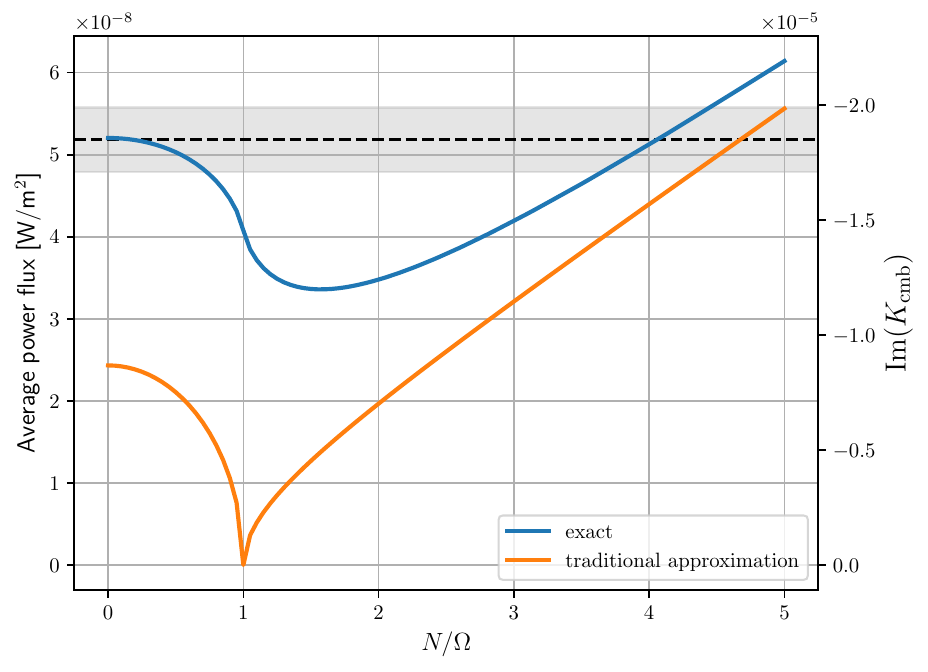}
    \caption{Average power flux at the CMB as a function of buoyancy for arbitrary topographic parameters ($h_\mathrm{rms}=6\,\mathrm{km}$, $k_0=2\pi/1000\,\mathrm{km}$, and $\mathcal{H}=0.7$). The orange and blue curves represent the solution with and without using the traditional approximation, respectively. The dashed line corresponds to the value of $\mathrm{Im}(K_\mathrm{cmb})$ inferred from the nutation data.}
    \label{fig:average_power_flux}
\end{figure}

Figure~\ref{fig:average_power_flux} shows the power flux averaged over the CMB surface (see \ref{sec:integratedpowerflux}) for arbitrary values of the topographic parameters. The axis on the left gives the value in $\mathrm{W/m^2}$ with the corresponding value of $\mathrm{Im}(K_\mathrm{cmb})$ given on the right axis for convenience. The horizontal dashed line and gray area are reproduced from Fig.~\ref{fig:Kcmb}. The two regimes are clearly visible.
Either inertial or gravity waves can in principle produce the power flux inferred from the nutation given a suitable topography. Note, however, that the model treats $N$ as a constant and does not take into account the finite depth of the stratified layer which is expected to greatly suppress the power flux \cite{Llewellyn-Young-2002:conversion}. Our results for $N>\Omega$ should therefore be treated as an upper bound and we conclude that the model favors $N\approx0$. 

\begin{figure}
    \centering
    \includegraphics[width=0.8\linewidth]{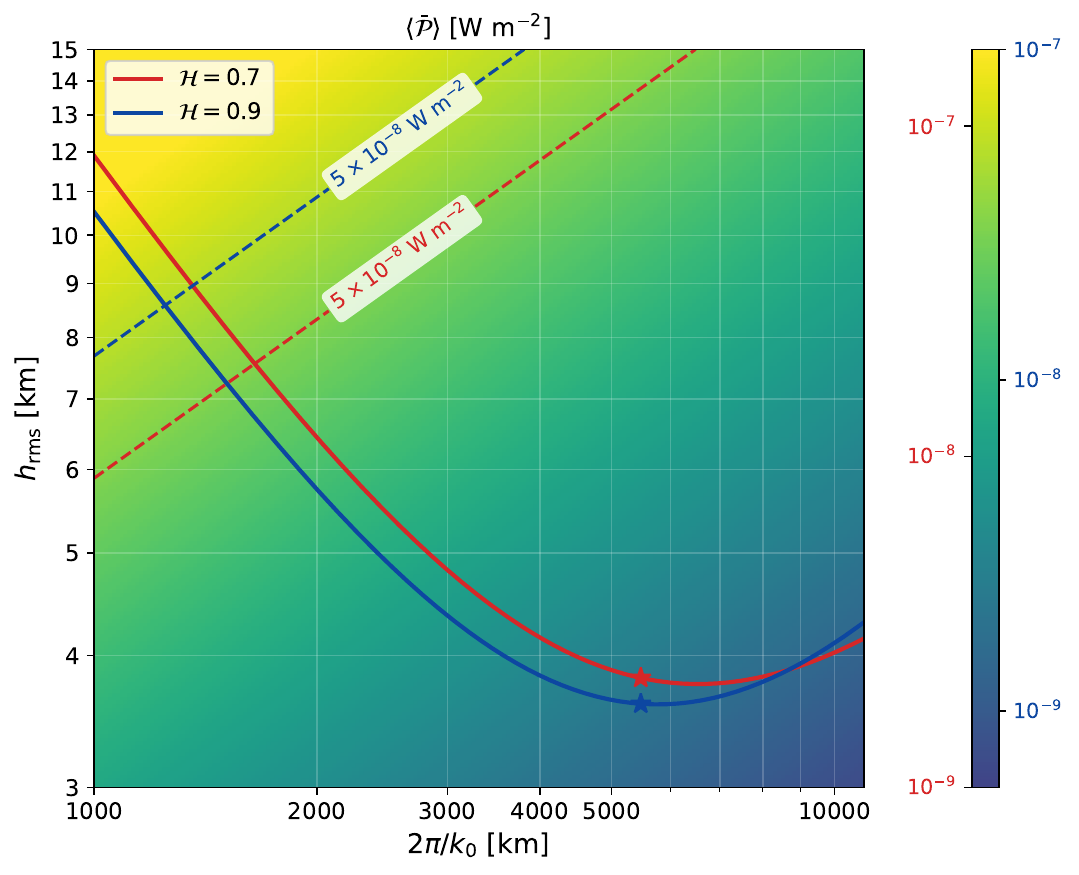}
    \caption{Trade-off between the amplitude of the topography ($h_\mathrm{rms}$) and the characteristic topographic wavelength ($2\pi/k_0$) for 2 values of the Hurst exponent ($\mathcal{H}$). The symbols correspond to the anchoring point obtained by setting $h_\star=2.5~\mathrm{km}$, and $k_\star=4/R$ in eq.~\eqref{eq:hrms_calibration}. Changing $\mathcal{H}$ rescales the color map (see main text). The two dashed lines correspond to the average power flux inferred from the nutation for both values of $\mathcal{H}$ considered.}
    \label{fig:trade_off_map}
\end{figure}

Assuming a well-mixed upper core and setting $N=0$, upon numerically integrating for the average power flux at the CMB, we find:
\begin{equation}
    \langle \bar{\mathcal{P}} \rangle\approx\frac{3\rho|\omega|^2R^2\Omega}{16\pi^{5/2}}\,\frac{\Gamma(\mathcal{H}-1/2)}{\Gamma(\mathcal{H})}\,h_\mathrm{rms}^2k_0,
\end{equation}
where we have rewritten the linear velocity of the tidal flow in terms of the angular velocity of rotation: $U_0=|\omega|R$. The model presents a trade-off between the values of $h_\mathrm{rms}$, $k_0$, and $\mathcal{H}$, illustrated on Figure~\ref{fig:trade_off_map}.
The simple power dependence in $h_\mathrm{rms}$ and $k_0$ is reflected in the straight contour lines of constant $\langle\bar{\mathcal{P}}\rangle$ on the log-log plot. The dependence on $\mathcal{H}$ is more subtle. Given the relatively weak dependence of the results on this parameter, we have collapsed the plots for $\mathcal{H}=0.7$ and $\mathcal{H}=0.9$ into a single figure: the background color map applies to both cases, but the color bar carries two separate scales---one for each value of $\mathcal{H}$---so that the same color maps to a slightly different $\langle\hat{\mathcal{P}}\rangle$ depending on which scaling is read. The dashed lines correspond to the average power flux inferred from the nutation for either cases. The star symbols correspond to the reference topography described by eq.~\eqref{eq:hrms_calibration}, with $k_0=k_\star=4/R$ and $h_\star=2.5\,\mathrm{km}$. The associated colored curves show the effect of varying $k_0$ while keeping the anchoring point and the corresponding value of $h_\mathrm{rms}$ fixed.

\begin{figure}
    \centering
    \includegraphics[width=0.8\linewidth]{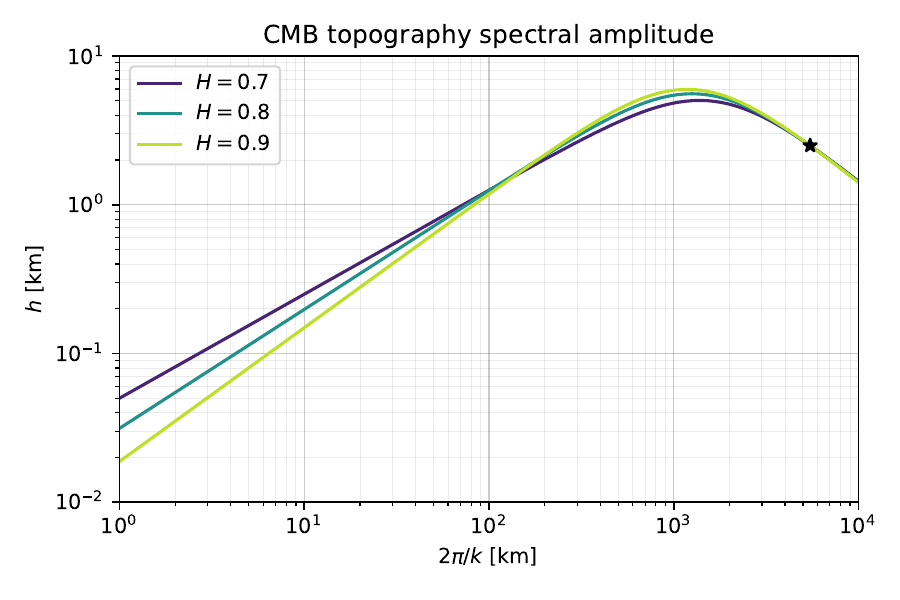}
    \caption{Spectral amplitude $h(k) = \sqrt{2\pi k^2\mathcal{S}(k)}$ of the CMB topography required to account entirely for the observed nutation dissipation, shown for three values of the Hurst exponent $\mathcal{H}$, with $h$ anchored to $h_\star = 2.5\,\mathrm{km}$ at $k_\star = 4/R$ (star symbol). The peak occurs at a wavelength of approximately $1500\,\mathrm{km}$ with an amplitude of $5$--$6\,\mathrm{km}$.}
    \label{fig:h_lambda}
\end{figure}

In order to fully account for the observed dissipation, the spectral amplitude $h(k)$ must peak at a wavelength of approximately $1500\,\mathrm{km}$. Figure~\ref{fig:h_lambda} shows $h(k)$ for three values of $\mathcal{H}$, with $h_\mathrm{rms}$ determined in each case by inverting eq.~\eqref{eq:hk} at the anchoring point $(k_\star, h_\star)$ with $k_\star= 4/R$ and $h_\star = 2.5\,\mathrm{km}$. The peak amplitude reaches $5$--$6\,\mathrm{km}$ and is only weakly sensitive to $\mathcal{H}$.  


\section{Discussion}

We computed the damping of the Earth's fluid core nutation due to the topography of the CMB. The dissipation results from the excitation of internal waves that remove energy and momentum from the tidal flow. Both inertial and gravity waves can contribute, with the latter requiring a stably stratified region at the top of the core. The dissipation is highest when the upper core is neutrally buoyant, allowing inertial waves to be excited over a broader portion of the CMB surface. This mechanism is presented as an alternative to the conventional electromagnetic coupling model, which we briefly reviewed and found insufficient to account for the full damping given current knowledge of the lower-mantle conductivity and CMB magnetic field.

The amount of dissipation depends on the shape of the topography at all length scales. We used a generic model of surface topography commonly applied to natural surfaces as well as to randomly generated surfaces for industrial applications, and identified a trade-off between the mean square amplitude of the topography and its characteristic wavenumber. To constrain this trade-off, we anchored the amplitude of the spectrum at large scales using the limited seismic data available at spherical harmonic degree $\ell=4$, and fitted its shape by matching the computed dissipation to that observed in the nutation.

Our results paint the picture of a CMB topography dominated by features at regional scales of order $1500\,\mathrm{km}$, with typical amplitudes of about $5\,\mathrm{km}$. Whether such a topography is realistic is difficult to assess with certainty: seismic constraints at these scales remain sparse and inconsistent across studies, with regional analyses reporting kilometer-scale features while global studies typically infer smaller amplitudes at the largest scales (see \citeA{Koelemeijer2021} for a review). The CMB topography need not be the sole contributor to the nutation dissipation, and a hybrid model combining topographic drag with electromagnetic coupling remains possible.

A second discrepancy reported in the literature concerns the amplitude of CMB topography at degree $\ell=2$, which seismic studies estimate to be roughly an order of magnitude larger than what nutation observations suggest. The $\ell=2$ component corresponds to the polar flattening of the fluid core, encoded in the dynamical flattening $e_f$ that appears in the FCN frequency (equation~\ref{eq:omegaFCN}). Nutation measurements thus constrain $e_f$, but only indirectly: the actually measured quantity is the combination $e_f + \mathrm{Re}(K_\mathrm{cmb})$, where $\mathrm{Re}(K_\mathrm{cmb})$ is the non-dissipative part of the CMB torque. Extracting $e_f$ therefore requires an assumption about the mechanism responsible for the dissipation.

\add{An important question concerns the fate of the energy radiated into the internal waves. Neither viscous nor Ohmic diffusion is efficient at the scales at which the waves are generated (see Appendix E), so the topography may sustain a persistent field of internal waves in the core interior that decays only very slowly. The core being a bounded domain, these waves return to the boundary after a short time rather than escaping. One may therefore ask whether the mechanism described here truly constitutes a sink of energy. This is a question of intrinsic interest for core energetics, but it is not the one on which the present constraint depends. The FCN is damped when energy is removed from the global, uniform vorticity flow that carries its kinetic energy (equation 2); once that energy resides in small-scale internal waves it has left the mode, irrespective of when it is ultimately converted to heat. What matters instead is whether waves returning to the boundary can do negative work and cancel the drag computed here.}

\add{The statistics of the topographic boundary make such a cancellation unlikely. All work exchanged at the CMB is mediated by the impermeability condition $w|_{z=0} = \mathbf{U}\cdot\nabla h$ of equation (7), so a returning wave can offset the pressure drag only if its pressure field is correlated with the topographic slope where it arrives. A wave carries the imprint of the topography that generated it, and returns at a location separated from its origin by a distance of order the core diameter, several times the characteristic scale of the features that dominate the dissipation ($\sim 1500$\,km, Figure 6). Generation and arrival patches are thus effectively independent realizations of the same random surface, and the ensemble-averaged correlation between the returning pressure field and the local slope vanishes. Returning waves therefore contribute a fluctuating force with zero mean rather than a systematic cancellation, and the secular energy transfer computed in equation (21) is unaffected to leading order.}

If electromagnetic coupling is taken to account for the full dissipation, then $\mathrm{Re}(K_\mathrm{cmb})\approx3\times10^{-5}$ \cite{BuffettEtAl2002,MathewsGuo2005}, leaving an $e_f$ that exceeds the hydrostatic value by about $3.6\%$ \cite{ZhangShen2021a}. The topographic coupling model presented here, by contrast, predicts a purely dissipative torque, $\mathrm{Re}(K_\mathrm{cmb})=0$, yielding an excess in $e_f$ of about $4.7\%$ compared to hydrostatic. The corresponding $\ell=2$ topographic amplitude inferred from nutation is therefore larger under our model than under the electromagnetic scenario, bringing it closer to the seismic estimates.

The advantage of the topographic power spectrum used in this study is its simplicity, and a natural extension of the present work would be to explore how variations in its shape affect our results. Given the strong dependence of the internal wave intensity on direction of propagation, relaxing the assumption of isotropy would be a particularly interesting exercise. The theory presented here could also be adapted to investigate the effect of CMB topography on length-of-day variations. In that context, \citeA{GlaneBuffett2018,MonvilleEtAl2025} have shown how topography, buoyancy, and magnetic field can interact non-trivially to enhance the coupling, with stronger stratification leading to stronger coupling at decadal timescales --- the opposite of what we find for nutation, where coupling is maximal under neutral buoyancy. 
\add{
Tyler and Kuang (2014) similarly found that a sufficiently stratified layer may be driven into resonance by the semidiurnal tide, at rates well above those associated with the nutations.
}
Reconciling nutation, length-of-day, and seismological constraints within a single picture of the CMB is an interesting question for future work.

\begin{appendix}
\section{Derivation of the power flux formula}
\label{sec:derivation}

Following \citeA{Bell1975}, we introduce the horizontal coordinates vector in a moving frame where the base flow appears at rest:
\begin{equation}
    \boldsymbol{\xi}\equiv\mathbf{x}-\int_0^t\mathbf{U}(\tau)d\tau,
    \label{eq:xi}
\end{equation}
where $\mathbf{x}^\mathsf{T}=(x,y)$ is the horizontal position vector in the rest frame. The vertical velocity may be decomposed into Fourier modes in either frame:
\begin{equation}
    w=\frac{1}{4\pi^2}\int \hat{w}~e^{i\mathbf{k}\cdot\mathbf{x}}~d^2k=\frac{1}{4\pi^2}\int \tilde{w}~e^{i\mathbf{k}\cdot\boldsymbol{\xi}}~d^2k,
    \label{eq:fourier}
\end{equation}
from which we deduce the relation between a Fourier component in the rest frame, $\hat{w}$, and in the moving frame, $\tilde{w}$:
\begin{equation}
    \hat{w}=\tilde{w}e^{-i\mathbf{k}\cdot\int_0^t\mathbf{U}(\tau)d\tau}.
\end{equation}
Injecting into the boundary condition \eqref{eq:bc} yields:
\begin{align}
    w|_{z=0}&=\mathbf{U}\cdot\mathbf{\nabla}h\nonumber\\
    \leftrightarrow\tilde{w}_0&=i(\mathbf{k}\cdot\mathbf{U})\hat{h}e^{i\mathbf{k}\cdot\int_0^t\mathbf{U}(\tau)d\tau}\nonumber\\
    \leftrightarrow\tilde{w}_0&=\hat{h}~\frac{d}{dt}e^{i\mathbf{k}\cdot\int_0^t\mathbf{U}(\tau)d\tau},
    \label{eq:bc_fourier}
\end{align}
where $\hat{w}_0\equiv\hat{w}_{z=0}$, and likewise for $\tilde{w}$. Using the base flow of equation~\eqref{eq:baseflow}, we compute the argument of the exponential in equation~\eqref{eq:bc_fourier}:
\begin{align}
    \mathbf{k}\cdot\int_0^t\mathbf{U}(\tau)d\tau&=\frac{k_xU_x}{\omega_0}\left(\sin{(\omega_0t+\varphi)}-\sin\varphi\right)+\frac{k_yU_y}{\omega_0}\left(\cos{(\omega_0t+\varphi)-\cos\varphi}\right)\nonumber\\
    &=\beta\sin{(\omega_0t+\varphi+\gamma)}-\beta\sin{(\varphi+\gamma)},
\end{align}
where we have defined:
\begin{align}
    \beta=\sqrt{\frac{k_x^2U_x^2+k_y^2U_y^2}{\omega_0^2}},&&\gamma=\arctan{\left(\frac{k_yU_y}{k_xU_x}\right)}.
    \label{eq:betagammadef}
\end{align}
It is useful to define the polar coordinates in Fourier space:\begin{equation}
\begin{cases}
    k_x=k\cos\alpha,\\
    k_y=k\sin\alpha.
    \label{eq:polar}
\end{cases}
\end{equation}
Following \citeA{Bell1975} again, we apply the \textit{Jacobi-Anger} relation to the boundary condition and find:
\begin{equation}
    \tilde{w}_0=\hat{h}\sum_{n=-\infty}^\infty J_n(\beta)(in\omega_0)e^{in\omega_0t}e^{in(\varphi+\gamma)}e^{-i\beta\sin(\varphi+\gamma)}.
    \label{eq:bc_JA}
\end{equation}
From equation \eqref{eq:bc_JA}, we can guess that the oscillation of the tidal flow over the topography at the frequency $\omega_0$ generates waves at the same frequency as well as higher harmonics. 

Equations~\eqref{eq:momx} to \eqref{eq:incompressibility} can be combined into a single equation for $w$:
\begin{equation}
    \left(\frac{\mathrm{D}^2}{\mathrm{D} t^2}\nabla^2+(\mathbf{f}\cdot\mathbf{\nabla})^2+N^2\nabla_h^2\right)w=0,
    \label{eq:eqw}
\end{equation}
where $\mathbf{f}^\mathsf{T}=(0,\tilde{f},f)$, and $\nabla_h^2=\left(\frac{\partial^2}{\partial x^2}+\frac{\partial^2}{\partial y^2}\right)$. Based on \eqref{eq:bc_JA}, we expect the solution to be a superposition of harmonic functions of the form:
\begin{equation}
    \tilde{w}=\sum_{n=-\infty}^{\infty}\tilde{w}^{(n)}e^{in\omega_0 t}.
\end{equation}
In Fourier space, equation \eqref{eq:eqw} applied to individual harmonics becomes:
\begin{equation}
    \left(\underbrace{(f^2-n^2\omega_0^2)}_{A}\frac{d^2}{dz^2}+\underbrace{2ikff_s}_{2iB}\frac{d}{dz}+\underbrace{(n^2\omega_0^2-N^2-f_s^2)k^2}_{C}\right)\tilde{w}^{(n)}=0,
\end{equation}
where we have used \eqref{eq:polar}, and we have defined $f_s=\tilde{f}\sin{\alpha}$. Inspired by \citeA{GerkemaZimmerman2008}, we define the ansatz:
\begin{equation}
    \tilde{w}^{(n)}=\tilde{w}^{(n)}_0e^{imz},
    \label{eq:solw}
\end{equation}
where the vertical wave number is defined as:
\begin{align}
    m&=\pm\sqrt{\frac{CA+B^2}{A^2}}-\frac{B}{A}\nonumber\\
    \leftrightarrow m&=\frac{k}{(f^2-n^2\omega_0^2)}\left(-\mathrm{sgn}(n)\sqrt{(n^2\omega_0^2-N^2)(f^2-n^2\omega_0^2)+n^2\omega_0^2f_s^2}-ff_s\right)
\end{align}
where we have inserted $-\mathrm{sgn(n)}$ to ensure downward energy propagation \cite{Bell1975a}.

The next step is to find expressions for the horizontal components of the velocity, and pressure. These will have a plane wave form analogous to equation \eqref{eq:solw} and can be recovered from equations~\eqref{eq:momx} to \eqref{eq:incompressibility} in Fourier space:
\begin{subequations}
\begin{align}
in\omega_0 \tilde{u}^{(n)}_0-f \tilde{v}^{(n)}_0+\tilde{f} \tilde{w}^{(n)}_0 & =-\frac{i k}{\rho} \cos \alpha \tilde{p}^{(n)}_0 \\
in\omega_0 \tilde{v}^{(n)}_0+f \tilde{u}^{(n)}_0 & =-\frac{i k}{\rho} \sin \alpha \tilde{p}^{(n)}_0 \\
in\omega_0 \tilde{w}^{(n)}_0-\tilde{f} \tilde{v}^{(n)}_0 & =-\frac{i}{\rho}m \tilde{p}^{(n)}_0+\tilde{b}^{(n)}_0\\
in\omega_0\tilde{b}^{(n)}_0+N^2\tilde{w}^{(n)}_0&=0\\
ik(\tilde{u}^{(n)}_0\cos{\alpha}+\tilde{v}^{(n)}_0\sin{\alpha})&=-im\tilde{w}^{(n)}_0,
\end{align}
\end{subequations}
from which the pressure coefficient proves to be:
\begin{align}
    \tilde{p}^{(n)}_0&=\frac{\rho}{n\omega_0 k^2}\left(n^2\omega_0^2-f^2\right)m \tilde{w}^{(n)}_0-\frac{\rho f f_s}{n\omega_0 k} \tilde{w}^{(n)}_0+i \frac{\rho f_c}{k} \tilde{w}^{(n)}_0\nonumber\\
    &=\mathrm{sgn(n)}\frac{\rho}{n\omega_0 k}\sqrt{(n^2\omega_0^2-N^2)(f^2-n^2\omega_0^2)+n^2\omega_0^2f_s^2}~\tilde{w}^{(n)}_0+i \frac{\rho f_c}{k} \tilde{w}^{(n)}_0
    \label{eq:solp}
\end{align}
where $f_c=\tilde{f}\cos{\alpha}$.

We get the expression for the power flux by means of equation~\eqref{eq:powerfluxdef}, which in the moving frame reads:
\begin{equation}
    \mathcal{P}=\frac{1}{A}\int_Ap^*w|_{z=0}~d^2\xi.
\end{equation}
The average power flux over the fundamental period is then:
\begin{align}
    \bar{\mathcal{P}}&=\frac{\omega_0}{2\pi}\int_0^{2\pi/\omega_0}\frac{1}{A}\int_Ad^2\xi~p^*w|_{z=0}~d^2\xi~dt\nonumber\\
    &=\frac{\omega_0}{2\pi}\frac{1}{4\pi^2}\int_0^{2\pi/\omega_0}\frac{1}{A}\int \frac{1}{2}(\tilde{p}_0^*\tilde{w}_0+\tilde{p}_0\tilde{w}^*_0)~d^2k~dt.
\end{align}
Focusing on the second term in the parentheses and using equations~\eqref{eq:bc_JA} and \eqref{eq:solp}:
\begin{multline}
  \int_0^{2\pi/\omega_0}\tilde{w}_0^*\tilde{p}_0~dt= 
  \sum_{n=-\infty}^\infty\sum_{s=-\infty}^\infty 
  \frac{\rho|\hat{h}|^2}{k}(s\omega_0)
  \left(\sqrt{(s^2\omega_0^2 - N^2)(f^2 - s^2\omega_0^2) + s^2\omega_0^2 f_s^2}+is\omega_0f_c\right) \\
  \times \mathrm{sgn(s)}J_n^*(\beta) J_s(\beta) e^{i(s-n)\gamma}
  \underbrace{\int_0^{2\pi/\omega_0}e^{i(s-n)\omega_0t}~dt}_{(2\pi/\omega_0)\delta_{n,s}}.
\end{multline}
Adding up the other term in the parentheses, we finally obtain:
\begin{equation}
    \bar{\mathcal{P}}=\frac{\rho}{2\pi^2}\sum_{n=1}^{\infty}(n\omega_0)\frac{1}{A}\int \frac{|\hat{h}|^2}{k}J_n(\beta)^2\sqrt{(n^2\omega_0^2 - N^2)(f^2 - n^2\omega_0^2) + n^2\omega_0^2 f_s^2}~d^2k,
    \label{eq:powerfluxapp}
\end{equation}
which is equivalent to equation~\eqref{eq:powerflux}, and which generalizes the expression found by \cite{Bell1975} under the traditional approximation.

\section{Quasi-static approximation}
\label{sec:QSapprox}

If $\beta\sim Uk/\omega_0$ is small, the terms for which $n=\pm1$ dominate the sum in equation~\eqref{eq:powerflux} (or equivalently equation~\ref{eq:powerfluxapp}). This is the so called \textit{quasi-static} approximation, under which the Bessel functions can be approximated as $J_n(\beta)\approx(\beta/2)^n/n!$. The power flux then approximates to:
\begin{equation}
    \bar{\mathcal{P}}\approx\frac{\rho}{8\pi^2|\omega_0|}\int_0^\infty \mathcal{S}(k)~k^2 ~dk\int_0^{2\pi}(U_x^2\cos^2\alpha+U_y^2\sin^2\alpha)\sqrt{(\omega_0^2 - N^2)(f^2 - \omega_0^2) + \omega_0^2 f_s^2}~d\alpha,\label{eq:powerfluxQS}
\end{equation}
where we have used the polar coordinates defined in equation~\eqref{eq:polar}, and we have assumed $\mathcal{S}(k_x,k_y)=\mathcal{S}(k)$ (isotropic). Under these hypotheses, the integral over $k$ can be computed independently. For the topographic model of equation~\eqref{eq:powerspectrum}, this gives:
\begin{equation}
    \int_0^\infty \mathcal{S}(k)~k^2 ~dk=\frac{h_\mathrm{rms}^2k_0}{4\sqrt{\pi}}\frac{\Gamma(\mathcal{H}-1/2)}{\Gamma(\mathcal{H})},
    \label{eq:intk^2S(k)}
\end{equation}
where $\Gamma$ is the Euler Gamma function \cite{NIST:DLMF}. In general, the remaining $\alpha$ integral must computed numerically. It becomes analytical when the traditional approximation is invoked.

\section{Traditional approximation}

The traditional approximation neglects the terms proportional to $\tilde{f}$ in equations~\eqref{eq:momx} and \eqref{eq:momz}. This amounts to setting $f_s=0$ in equation~\eqref{eq:powerfluxQS}. The integral over $\alpha$ may then be carried out analytically, which yields (using equation~\ref{eq:intk^2S(k)}):
\begin{align}
    \bar{\mathcal{P}}_\mathrm{TA}&=\frac{\rho}{4\pi|\omega_0|}(U_x^2+U_y^2)\sqrt{(\omega_0^2 - N^2)(f^2 - \omega_0^2)}\int_0^\infty k^2~\mathcal{S}(k)~dk\\
    &=\frac{\rho h_\mathrm{rms}^2k_0}{16|\omega_0|\sqrt{\pi^3}}\frac{\Gamma(\mathcal{H}-1/2)}{\Gamma(\mathcal{H})}(U_x^2+U_y^2)\sqrt{(\omega_0^2 - N^2)(f^2 - \omega_0^2)}.
    \label{eq:powerfluxQSTA}
\end{align}
This expression for the power flux depends on the latitude through $f$, as well as $U_x$, and $U_y$, read from equation~\eqref{eq:baseflow}.

\section{Power flux averaged over the CMB}
\label{sec:integratedpowerflux}
We compute the power flux averaged over the CMB surface by integrating over latitudes and longitudes. Since the power flux does not depend on the longitude, the final formula reads:
\begin{equation}
    \langle\mathcal{\bar{P}}\rangle=\frac{1}{2}\int_{-\pi/2}^{\pi/2}\bar{\mathcal{P}}\cos\vartheta~d\vartheta.
\end{equation}
We evaluate this integral numerically.

\section{Fate of the radiated energy}
\label{sec:damping}
\add{
The damping of the FCN is set by the work done at the boundary, equation (21),
and does not depend on where the radiated energy is eventually converted to
heat. That is a separate question, and one the topographic spectrum is well
suited to address, since it fixes not only the total power radiated but also its
distribution across wavelengths.

Each Fourier component of the topography radiates waves of the same horizontal
wavelength, and the power it delivers follows directly from the spectrum. Since
$h(k)^2 = 2\pi k^2 \mathcal{S}(k)$, the integrand of equation (B1) is
proportional to $h(k)^2$, so Figure 6 may be read as the spectrum of radiated
power as well as of topographic amplitude. Waves are excited across the whole
range of scales it spans. The distribution is broad: although $h(k)$ peaks near
$1500$\,km, it decays only as $k^{-\mathcal{H}}$ beyond that, so half of the
total power is radiated at wavelengths below roughly $200$\,km for
$\mathcal{H} = 0.7$, or $500$\,km for $\mathcal{H} = 0.9$. The same slow decay
is what requires $\mathcal{H} > 1/2$ for the power flux to converge.

The fate of a wave depends strongly on its wavelength, because both the rate at
which it decays and the time it spends in the core do. Diffusive damping rates
grow as $k^2$, while the group velocity of inertial waves at the forcing
frequency is $\sqrt{3}\,\Omega/|\mathbf{k}|$ (Vallis, 2017), where
$|\mathbf{k}|$ is the total wavenumber, which exceeds the topographic wavenumber
$k$ by a factor of order unity; taking $|\mathbf{k}| \approx 1.3\,k$ gives
$c_g \approx 1.4\,\Omega/k$, so short waves decay faster and travel more slowly
between encounters with a boundary. A wave crossing the fluid core covers a
distance of between $2300$\,km, the thickness of the shell, and $6500$\,km for a
path grazing the inner core; we take a typical value $D \approx 4000$\,km, of
the order of the core radius. At the peak of the spectrum this gives
$c_g \approx 24\,\mathrm{m\,s^{-1}}$ and a travel time of about two days, but
the latter grows in proportion to $k$ and reaches several months at kilometric
wavelengths. The energy lost between encounters therefore scales as $k^3$, and
the spectrum separates into waves absorbed in the interior and waves that return
to the boundary essentially intact. Equating the Ohmic damping rate
$k^2/(\mu_0\sigma_\mathrm{f})$ to the inverse travel time $1.4\,\Omega/(Dk)$
places the division at a wavelength of $25$\,km. This is an upper bound, as it
assumes the magnetic perturbation carries a substantial share of the wave
energy. Where the field is weak, in the sense that the Alfv\'en frequency
$kV_\mathrm{A}$ is small compared with $\Omega$, the induced field satisfies
$b \sim (kV_\mathrm{A}/\Omega)\sqrt{\mu_0\rho}\,u$, so only a fraction
$(kV_\mathrm{A}/\Omega)^2$ of the wave energy is magnetic and available to Ohmic
decay. The same balance then gives $2-3$\,km for internal field strengths of
$1-10$\,mT. This is also the scale at which $kV_\mathrm{A}$ becomes comparable
to $\Omega$ and the waves acquire a magnetic character, so the two estimates
converge from either side. Viscosity alone would require wavelengths shorter
than $200$\,m. Geodynamo simulations operating at the physical conditions of the
core independently place the characteristic length of Ohmic dissipation at a few
tens of kilometres (Aubert, 2026), obtained from the ratio of the magnetic
energy to that of its gradient rather than from any propagation argument, in
agreement with the upper estimate above.

Part of the radiated energy is therefore dissipated directly, without any
transfer between scales: taking $3$\,km as the transition, this amounts to
$1-9\%$ of the total for $\mathcal{H}$ between $0.9$ and $0.7$, rising to
$5-22\%$ under the more optimistic estimate. The remainder, radiated at longer
wavelengths, must reach smaller scales before it can be converted to heat, a
reduction of some two decades from the median injection scale. Reflection from
the boundary offers a natural route, since the topographic spectrum retains
power at all wavenumbers and each encounter scatters energy into shorter waves;
the long travel times of the waves concerned, months rather than days, leave
ample opportunity for this to occur. We do not attempt to quantify the
efficiency of the transfer. We note that the viscous scale quoted above lies
within the range of small-scale topography considered by Le Mou\"el et al.
(2006), suggesting that the two mechanisms operate at opposite ends of the same
spectrum rather than as alternatives.}

\end{appendix}

%



%
%

\section*{Data Availability Statement}
The \textsc{Jupyter} notebooks used to generate Figures~2 to 6 have been made publicly available at \cite{jeremy_rekier_2026_20021948}.

\section*{Conflict of Interest declaration}
The authors declare there are no conflicts of interest for this manuscript.

\acknowledgments
The research leading to these results has received funding from the European Research Council (ERC) under the European Union's Horizon 2020 research and innovation program (Synergy Grant 855677 GRACEFUL). D.A. acknowledges support from the U.S. Department of Energy, Office of Science, Office of Advanced Scientific Computing Research, Department of Energy Computational Science Graduate Fellowship Award Number (DE-SC0023112).

%
\bibliography{bibliography}

@article{Bell1975,
  title = {Topographically Generated Internal Waves in the Open Ocean},
  author = {Bell, T. H.},
  year = {1975},
  month = jan,
  journal = {Journal of Geophysical Research},
  volume = {80},
  number = {3},
  pages = {320--327},
  issn = {01480227},
  doi = {10.1029/JC080i003p00320},
  urldate = {2024-04-23},
  copyright = {http://doi.wiley.com/10.1002/tdm\_license\_1.1},
  langid = {english}
}

@article{Bell1975a,
  title = {Lee Waves in Stratified Flows with Simple Harmonic Time Dependence},
  author = {Bell, T. H.},
  year = {1975},
  month = feb,
  journal = {Journal of Fluid Mechanics},
  volume = {67},
  pages = {705--722},
  issn = {0022-1120},
  doi = {10.1017/S0022112075000560},
  urldate = {2024-05-15}
}

@article{Buffett2010a,
  title = {Chemical Stratification at the Top of {{Earth}}'s Core: {{Constraints}} from Observations of Nutations},
  shorttitle = {Chemical Stratification at the Top of {{Earth}}'s Core},
  author = {Buffett, Bruce A.},
  year = {2010},
  month = aug,
  journal = {Earth and Planetary Science Letters},
  volume = {296},
  number = {3-4},
  pages = {367--372},
  issn = {0012821X},
  doi = {10.1016/j.epsl.2010.05.020},
  urldate = {2024-10-21},
  copyright = {https://www.elsevier.com/tdm/userlicense/1.0/},
  langid = {english}
}

@article{PerssonEtAl2004,
  title = {On the Nature of Surface Roughness with Application to Contact Mechanics, Sealing, Rubber Friction and Adhesion},
  author = {Persson, B. N. J. and Albohr, O. and Tartaglino, U. and Volokitin, A. I. and Tosatti, E.},
  year = {2004},
  month = dec,
  journal = {Journal of Physics: Condensed Matter},
  volume = {17},
  number = {1},
  pages = {R1},
  issn = {0953-8984},
  doi = {10.1088/0953-8984/17/1/R01},
  urldate = {2024-10-21},
  langid = {english}
}

@article{CandelaEtAl2012,
  title = {Roughness of Fault Surfaces over Nine Decades of Length Scales},
  author = {Candela, Thibault and Renard, Fran{\c c}ois and Klinger, Yann and Mair, Karen and Schmittbuhl, Jean and Brodsky, Emily E.},
  year = {2012},
  month = aug,
  journal = {Journal of Geophysical Research: Solid Earth},
  volume = {117},
  number = {B8},
  pages = {2011JB009041},
  issn = {0148-0227},
  doi = {10.1029/2011JB009041},
  urldate = {2024-12-02},
  copyright = {http://onlinelibrary.wiley.com/termsAndConditions\#vor},
  langid = {english}
}

@article{GagnonEtAl2006,
  title = {Multifractal Earth Topography},
  author = {Gagnon, J.-S. and Lovejoy, S. and Schertzer, D.},
  year = {2006},
  month = oct,
  journal = {Nonlinear Processes in Geophysics},
  volume = {13},
  number = {5},
  pages = {541--570},
  issn = {1607-7946},
  doi = {10.5194/npg-13-541-2006},
  urldate = {2024-09-13},
  copyright = {https://creativecommons.org/licenses/by-nc-sa/2.5/},
  langid = {english}
}

@article{VondrakRon2017,
  title = {New Method for Determining Free Core Nutation Parameters, Considering Geophysical Effects},
  author = {Vondr{\'a}k, J. and Ron, C.},
  year = {2017},
  month = aug,
  journal = {Astronomy \& Astrophysics},
  volume = {604},
  pages = {A56},
  issn = {0004-6361, 1432-0746},
  doi = {10.1051/0004-6361/201730635},
  urldate = {2024-12-02},
  langid = {english}
}

@article{ZhuEtAl2021,
  title = {Quantification of Corrections for the Main Lunisolar Nutation Components and Analysis of the Free Core Nutation from {{VLBI-observed}} Nutation Residuals},
  author = {Zhu, Ping and Triana, Santiago A. and Rekier, J{\'e}r{\'e}my and Trinh, Antony and Dehant, V{\'e}ronique},
  year = {2021},
  journal = {Journal of Geodesy},
  volume = {123},
  publisher = {Springer Berlin Heidelberg},
  issn = {1432-1394},
  doi = {10.1007/s00190-021-01513-9},
  copyright = {All rights reserved},
  isbn = {0123456789}
}

@article{ZhangShen2021a,
  title = {Core--Mantle Topographic Coupling: A Parametric Approach and Implications for the Formulation of a Triaxial Three-Layered {{Earth}} Rotation},
  shorttitle = {Core--Mantle Topographic Coupling},
  author = {Zhang, Huifeng and Shen, Wenbin},
  year = {2021},
  month = mar,
  journal = {Geophysical Journal International},
  volume = {225},
  number = {3},
  pages = {2060--2074},
  issn = {0956-540X, 1365-246X},
  doi = {10.1093/gji/ggab079},
  urldate = {2024-12-02},
  copyright = {https://academic.oup.com/journals/pages/open\_access/funder\_policies/chorus/standard\_publication\_model},
  langid = {english}
}

@article{NurulHudaEtAl2020,
  title = {Nutation Terms Adjustment to {{VLBI}} and Implication for the {{Earth}} Rotation Resonance Parameters},
  author = {Nurul~Huda, I and Lambert, S and Bizouard, C and Ziegler, Y},
  year = {2020},
  month = feb,
  journal = {Geophysical Journal International},
  volume = {220},
  number = {2},
  pages = {759--767},
  issn = {0956-540X, 1365-246X},
  doi = {10.1093/gji/ggz468},
  urldate = {2024-12-02},
  copyright = {https://academic.oup.com/journals/pages/open\_access/funder\_policies/chorus/standard\_publication\_model},
  langid = {english}
}

@article{BuffettEtAl2002,
  title = {Modeling of Nutation and Precession: {{Effects}} of Electromagnetic Coupling},
  shorttitle = {Modeling of Nutation and Precession},
  author = {Buffett, B. A. and Mathews, P. M. and Herring, T. A.},
  year = {2002},
  month = apr,
  journal = {Journal of Geophysical Research: Solid Earth},
  volume = {107},
  number = {B4},
  pages = {ETG 5-1-ETG 5-14},
  issn = {01480227},
  doi = {10.1029/2000JB000056},
  urldate = {2022-03-31},
  langid = {english}
}

@article{KootEtAl2010,
  title = {Constraints on the Coupling at the Core–Mantle and Inner Core Boundaries Inferred from Nutation Observations},
  author = {Koot, L. and Dumberry, M. and Rivoldini, A. and De Viron, O. and Dehant, V.},
  year = {2010},
  journal = {Geophysical Journal International},
  shortjournal = {Geophysical Journal International},
  volume = {182},
  number = {3},
  pages = {1279--1294},
  issn = {0956-540X},
  doi = {10.1111/j.1365-246X.2010.04711.x}
}

@article{Braginsky1993,
  title = {{{MAC-Oscillations}} of the {{Hidden Ocean}} of the {{Core}}},
  author = {Braginsky, S. I.},
  year = {1993},
  journal = {Journal of geomagnetism and geoelectricity},
  volume = {45},
  number = {11-12},
  pages = {1517--1538},
  doi = {10.5636/jgg.45.1517}
}

@article{Braginsky1999,
  title = {Dynamics of the Stably Stratified Ocean at the Top of the Core},
  author = {Braginsky, Stanislav I.},
  year = {1999},
  month = feb,
  journal = {Physics of the Earth and Planetary Interiors},
  volume = {111},
  number = {1},
  pages = {21--34},
  issn = {0031-9201},
  doi = {10.1016/S0031-9201(98)00143-5}
}

@article{Helffrich2013,
  title = {Causes and Consequences of Outer Core Stratification},
  author = {Helffrich, George and Kaneshima, Satoshi},
  year = {2013},
  month = oct,
  journal = {Physics of the Earth and Planetary Interiors},
  volume = {223},
  pages = {2--7},
  issn = {00319201},
  doi = {10.1016/j.pepi.2013.07.005}
}

@article{Higgins1971,
  title = {The Adiabatic Gradient and the Melting Point Gradient in the Core of the {{Earth}}},
  author = {Higgins, G. and Kennedy, G. C.},
  year = {1971},
  journal = {Journal of Geophysical Research (1896-1977)},
  volume = {76},
  number = {8},
  pages = {1870--1878},
  issn = {2156-2202},
  doi = {10.1029/JB076i008p01870},
  copyright = {Copyright 1971 by the American Geophysical Union.}
}

@article{Kaneshima2018,
  title = {Array Analyses of {{SmKS}} Waves and the Stratification of {{Earth}}'s Outermost Core},
  author = {Kaneshima, Satoshi},
  year = {2018},
  month = mar,
  journal = {Physics of the Earth and Planetary Interiors},
  volume = {276},
  pages = {234--246},
  issn = {00319201},
  doi = {10.1016/j.pepi.2017.03.006}
}

@article{Whaler1980,
  title = {Does the Whole of the {{Earth}}'s Core Convect?},
  author = {Whaler, K. A.},
  year = {1980},
  month = oct,
  journal = {Nature},
  volume = {287},
  number = {5782},
  pages = {528--530},
  publisher = {Nature Publishing Group},
  issn = {1476-4687},
  doi = {10.1038/287528a0},
  copyright = {1980 Springer Nature Limited}
}

@article{Tanaka1993,
  title = {Velocities and {{Chemical Stratification}} in the {{Outermost Core}}},
  author = {Tanaka, Satoru and Hamaguchi, Hiroyuki},
  year = {1993},
  journal = {Journal of geomagnetism and geoelectricity},
  volume = {45},
  number = {11-12},
  pages = {1287--1301},
  doi = {10.5636/jgg.45.1287}
}

@article{Tang2015,
  title = {Seismological Evidence for a Non-Monotonic Velocity Gradient in the Topmost Outer Core},
  author = {Tang, Vivian and Zhao, Li and Hung, Shu-Huei},
  year = {2015},
  month = feb,
  journal = {Scientific Reports},
  volume = {5},
  number = {1},
  pages = {8613},
  publisher = {Nature Publishing Group},
  issn = {2045-2322},
  doi = {10.1038/srep08613},
  copyright = {2015 The Author(s)}
}

@book{DehantMathews2015,
  title = {Precession, {{Nutation}} and {{Wobble}} of the {{Earth}}},
  author = {Dehant, V. and Mathews, P. M.},
  year = {2015},
  publisher = {Cambridge University Press},
  address = {Cambridge},
  doi = {10.1017/CBO9781316136133},
  isbn = {978-1-107-09254-9}
}

@article{BarikAngappan2024,
  title = {{{planetMagFields}}: {{A Python}} Package for Analyzing Andplotting Planetary Magnetic Field Data},
  shorttitle = {{{planetMagFields}}},
  author = {Barik, Ankit and Angappan, Regupathi},
  year = {2024},
  month = may,
  journal = {Journal of Open Source Software},
  volume = {9},
  number = {97},
  pages = {6677},
  issn = {2475-9066},
  doi = {10.21105/joss.06677},
  urldate = {2025-01-24},
  copyright = {http://creativecommons.org/licenses/by/4.0/}
}

@article{MathewsGuo2005,
  title = {Viscoelectromagnetic Coupling in Precession-Nutation Theory},
  shorttitle = {Viscoelectromagnetic Coupling in Precession-Nutation Theory},
  author = {Mathews, P. M. and Guo, J. Y.},
  year = {2005},
  month = feb,
  journal = {Journal of Geophysical Research: Solid Earth},
  volume = {110},
  number = {B2},
  issn = {01480227},
  doi = {10.1029/2003JB002915},
  urldate = {2022-04-28},
  langid = {english}
}

@article{Buffett1992,
  title = {Constraints on Magnetic Energy and Mantle Conductivity from the Forced Nutations of the {{Earth}}},
  author = {Buffett, Bruce A.},
  year = {1992},
  journal = {Journal of Geophysical Research},
  volume = {97},
  number = {B13},
  pages = {19581},
  issn = {0148-0227},
  doi = {10.1029/92JB00977},
  urldate = {2022-03-31},
  langid = {english}
}

@article{gubbins_geomagnetic_2007,
    title = {Geomagnetic constraints on stratification at the top of {Earth}’s core},
    volume = {59},
    issn = {1880-5981},
    url = {https://doi.org/10.1186/BF03352728},
    doi = {10.1186/BF03352728},
    language = {en},
    number = {7},
    urldate = {2024-12-31},
    journal = {Earth, Planets and Space},
    author = {Gubbins, David},
    month = jul,
    year = {2007},
    pages = {661--664},
}

@article{buffett_geomagnetic_2014,
    title = {Geomagnetic fluctuations reveal stable stratification at the top of the {Earth}’s core},
    volume = {507},
    copyright = {2014 Springer Nature Limited},
    issn = {1476-4687},
    url = {https://www.nature.com/articles/nature13122},
    doi = {10.1038/nature13122},
    language = {en},
    number = {7493},
    urldate = {2024-12-31},
    journal = {Nature},
    author = {Buffett, Bruce},
    month = mar,
    year = {2014},
    note = {Publisher: Nature Publishing Group},
    pages = {484--487},
}

@article{lister_stratification_1998,
    title = {Stratification of the outer core at the core-mantle boundary},
    volume = {105},
    copyright = {https://www.elsevier.com/tdm/userlicense/1.0/},
    issn = {00319201},
    url = {https://linkinghub.elsevier.com/retrieve/pii/S0031920197000824},
    doi = {10.1016/S0031-9201(97)00082-4},
    language = {en},
    number = {1-2},
    urldate = {2024-12-31},
    journal = {Physics of the Earth and Planetary Interiors},
    author = {Lister, John R. and Buffett, Bruce A.},
    month = jan,
    year = {1998},
    pages = {5--19},
}

@article{Gubbins2013,
  title = {The Stratified Layer at the Core--Mantle Boundary Caused by Barodiffusion of Oxygen, Sulphur and Silicon},
  author = {Gubbins, D. and Davies, C. J.},
  year = {2013},
  month = feb,
  journal = {Physics of the Earth and Planetary Interiors},
  volume = {215},
  pages = {21--28},
  issn = {0031-9201},
  doi = {10.1016/j.pepi.2012.11.001}
}

@article{Gastine2020,
  title = {Dynamo-Based Limit to the Extent of a Stable Layer atop {{Earth}}'s Core},
  author = {Gastine, Thomas and Aubert, Julien and Fournier, Alexandre},
  year = {2020},
  month = aug,
  journal = {Geophysical Journal International},
  volume = {222},
  number = {2},
  pages = {1433--1448},
  issn = {0956-540X},
  doi = {10.1093/gji/ggaa250}
}

@article{Christensen2018,
  title = {Geodynamo Models with a Stable Layer and Heterogeneous Heat Flow at the Top of the Core},
  author = {Christensen, Ur},
  year = {2018},
  month = nov,
  journal = {Geophysical Journal International},
  volume = {215},
  number = {2},
  pages = {1338--1351},
  issn = {0956-540X, 1365-246X},
  doi = {10.1093/gji/ggy352}
}

@incollection{Koelemeijer2021,
  title = {Toward {{Consistent Seismological Models}} of the {{Core}}--{{Mantle Boundary Landscape}}},
  booktitle = {Geophysical {{Monograph Series}}},
  author = {Koelemeijer, Paula},
  editor = {Marquardt, Hauke and Ballmer, Maxim and Cottaar, Sanne and Konter, Jasper},
  year = {2021},
  month = jul,
  edition = {1},
  pages = {229--255},
  publisher = {Wiley},
  doi = {10.1002/9781119528609.ch9},
  copyright = {http://doi.wiley.com/10.1002/tdm\_license\_1.1},
  isbn = {978-1-119-52861-6 978-1-119-52860-9}
}

@article{Heyn2020,
  title = {Core-Mantle Boundary Topography and Its Relation to the Viscosity Structure of the Lowermost Mantle},
  author = {Heyn, Bj{\"o}rn H. and Conrad, Clinton P. and Tr{\o}nnes, Reidar G.},
  year = {2020},
  month = aug,
  journal = {Earth and Planetary Science Letters},
  volume = {543},
  pages = {116358},
  issn = {0012821X},
  doi = {10.1016/j.epsl.2020.116358}
}

@article{JacobsEtAl2017,
  title = {Quantitative Characterization of Surface Topography Using Spectral Analysis},
  author = {Jacobs, Tevis D B and Junge, Till and Pastewka, Lars},
  year = {2017},
  month = jan,
  journal = {Surface Topography: Metrology and Properties},
  volume = {5},
  number = {1},
  pages = {013001},
  issn = {2051-672X},
  doi = {10.1088/2051-672X/aa51f8},
  urldate = {2024-09-11},
  langid = {english}
}

@article{NaveiraGarabatoEtAl2013,
  title = {The {{Impact}} of {{Small-Scale Topography}} on the {{Dynamical Balance}} of the {{Ocean}}},
  author = {Naveira Garabato, Alberto C. and Nurser, A. J. George and Scott, Robert B. and Goff, John A.},
  year = {2013},
  month = mar,
  journal = {Journal of Physical Oceanography},
  volume = {43},
  number = {3},
  pages = {647--668},
  issn = {0022-3670, 1520-0485},
  doi = {10.1175/JPO-D-12-056.1},
  urldate = {2024-09-23},
  langid = {english}
}

@article{RodriguezEtAl2025,
  title = {On {{How}} to {{Determine Surface Roughness Power Spectra}}},
  author = {Rodriguez, N. and Gontard, L. and Ma, C. and Xu, R. and Persson, B. N. J.},
  year = {2025},
  month = mar,
  journal = {Tribology Letters},
  volume = {73},
  number = {1},
  pages = {18},
  issn = {1023-8883, 1573-2711},
  doi = {10.1007/s11249-024-01933-6},
  urldate = {2025-03-05},
  langid = {english}
}

@article{Llewellyn-Young-2002:conversion,
  title =	 {Conversion of the barotropic tide},
  author =	 {Llewellyn Smith, Stefan G and Young, WR},
  journal =	 {Journal of Physical Oceanography},
  volume =	 32,
  number =	 5,
  pages =	 {1554--1566},
  year =	 2002
}

@book{Vallis-2017:atmospheric,
  title =	 {Atmospheric and oceanic fluid dynamics},
  author =	 {Vallis, Geoffrey K},
  year =	 2017,
  publisher =	 {Cambridge University Press}
}

@software{jeremy_rekier_2026_20021948,
  author       = {Jérémy Rekier and
                  Ankit Barik},
  title        = {Topo Nutation (version 2.0) [Software]
                  },
  month        = may,
  year         = 2026,
  publisher    = {Zenodo},
  version      = {v2.0},
  doi          = {10.5281/zenodo.20021948},
  url          = {https://doi.org/10.5281/zenodo.20021948},
  swhid        = {swh:1:dir:3067d190da3aec85bd2fe41b3e88ecce0e2d9cc8
                   ;origin=https://doi.org/10.5281/zenodo.15748777;vi
                   sit=swh:1:snp:e01f861ae6d5901b86a1c34e9e206a2e5702
                   cbf1;anchor=swh:1:rel:a95ddc756d44ef51aaa16fb94854
                   08fab5d511fe;path=AnkitBarik-topo\_nutation-0362da5
                  },
}

@incollection{Olson2015,
  title = {Core {{Dynamics}}: {{An Introduction}} and {{Overview}}},
  shorttitle = {Core {{Dynamics}}},
  booktitle = {Treatise on {{Geophysics}}},
  author = {Olson, P.},
  year = {2015},
  pages = {1--25},
  publisher = {Elsevier},
  doi = {10.1016/B978-0-444-53802-4.00137-8},
  urldate = {2022-05-13},
  isbn = {978-0-444-53803-1},
  langid = {english}
}

@article{GlaneBuffett2018,
  title = {Enhanced {{Core-Mantle Coupling Due}} to {{Stratification}} at the {{Top}} of the {{Core}}},
  author = {Glane, Sebastian and Buffett, Bruce},
  year = {2018},
  month = oct,
  journal = {Frontiers in Earth Science},
  volume = {6},
  pages = {171},
  issn = {2296-6463},
  doi = {10.3389/feart.2018.00171},
  urldate = {2021-09-29},
  langid = {english}
}

@article{Aubert2025,
  title = {Rapid Geomagnetic Variations and Stable Stratification at the Top of {{Earth}}'s Core},
  author = {Aubert, Julien},
  year = {2025},
  month = may,
  journal = {Physics of the Earth and Planetary Interiors},
  volume = {362},
  pages = {107335},
  issn = {0031-9201},
  doi = {10.1016/j.pepi.2025.107335},
  urldate = {2025-06-03}
}

@article{MathewsEtAl2002,
  title = {Modeling of Nutation and Precession: {{New}} Nutation Series for Nonrigid {{Earth}} and Insights into the {{Earth}}'s Interior: {{NEW NUTATION SERIES AND THE EARTH}}'{{S INTERIOR}}},
  shorttitle = {Modeling of Nutation and Precession},
  author = {Mathews, P. M. and Herring, T. A. and Buffett, B. A.},
  year = {2002},
  month = apr,
  journal = {Journal of Geophysical Research: Solid Earth},
  volume = {107},
  number = {B4},
  pages = {ETG 3-1-ETG 3-26},
  issn = {01480227},
  doi = {10.1029/2001JB000390},
  urldate = {2021-10-15},
  langid = {english}
}

@article{Rekier2022a,
  title = {Free {{Core Nutation}} and {{Its Relation}} to the {{Spin-over Mode}}},
  author = {Rekier, J{\'e}r{\'e}my},
  year = {2022},
  month = jun,
  journal = {The Planetary Science Journal},
  volume = {3},
  number = {6},
  pages = {133},
  issn = {2632-3338},
  doi = {10.3847/PSJ/ac6ce2},
  urldate = {2022-06-07},
  copyright = {Creative Commons Attribution-NonCommercial-NoDerivatives 4.0 International License (CC-BY-NC-ND)},
  langid = {english}
}

@article{RekierEtAl2021,
  title = {Earth's {{Rotation}}: {{Observations}} and {{Relation}} to {{Deep Interior}}},
  shorttitle = {Earth's {{Rotation}}},
  author = {Rekier, J{\'e}r{\'e}my and Chao, Benjamin F. and Chen, Jianli and Dehant, V{\'e}ronique and Rosat, S{\'e}verine and Zhu, Ping},
  year = {2021},
  month = nov,
  journal = {Surveys in Geophysics},
  volume = {43},
  number = {1},
  pages = {149--175},
  issn = {1573-0956},
  doi = {10.1007/s10712-021-09669-x},
  urldate = {2023-01-11},
  copyright = {All rights reserved},
  langid = {english}
}

@article{ShihEtAl2023,
  title = {Turbulent Dissipation in the Boundary Layer of Precession-Driven Flow in a Sphere},
  author = {Shih, Sheng-An and Triana, Santiago Andr{\'e}s and Rekier, J{\'e}r{\'e}my and Dehant, V{\'e}ronique},
  year = {2023},
  month = jul,
  journal = {AIP Advances},
  volume = {13},
  number = {7},
  pages = {075025},
  issn = {2158-3226},
  doi = {10.1063/5.0146932},
  urldate = {2023-07-27},
  copyright = {All rights reserved}
}

@incollection{Tyburczy2015,
  title = {2.25 - Properties of Rocks and Minerals -- the Electrical Conductivity of Rocks, Minerals, and the Earth},
  booktitle = {Treatise on Geophysics (Second Edition)},
  author = {Tyburczy, J.A. and Du Frane, W.L.},
  editor = {Schubert, Gerald},
  year = {2015},
  edition = {Second Edition},
  pages = {661--672},
  publisher = {Elsevier},
  address = {Oxford},
  doi = {10.1016/B978-0-444-53802-4.00049-X},
  isbn = {978-0-444-53803-1}
}

@article{Mound2019,
  title = {Regional Stratification at the Top of {{Earth}}'s Core Due to Core--Mantle Boundary Heat Flux Variations},
  author = {Mound, Jon and Davies, Chris and Rost, Sebastian and Aurnou, Jon},
  year = {2019},
  month = jul,
  journal = {Nature Geoscience},
  volume = {12},
  number = {7},
  pages = {575--580},
  issn = {1752-0894, 1752-0908},
  doi = {10.1038/s41561-019-0381-z}
}

@misc{GerkemaZimmerman2008,
  title = {Lecture {{Notes}}: {{An}} Introduction to Internal Waves},
  author = {Gerkema, T and Zimmerman, J T F},
  year = {2008},
  publisher = {NIOZ Open Repository},
  langid = {english}
}

@article{MonvilleEtAl2025,
  title = {Topographic {{Drag}} at the {{Core-Mantle Interface}}},
  author = {Monville, R. and C{\'e}bron, D. and Jault, D.},
  year = 2025,
  journal = {Journal of Geophysical Research: Solid Earth},
  volume = {130},
  number = {4},
  pages = {e2024JB029770},
  issn = {2169-9356},
  doi = {10.1029/2024JB029770},
  urldate = {2025-12-05},
  copyright = {\copyright{} 2025. The Author(s).},
  langid = {english}
}

@article{HoEtAl2024,
  title = {Quantum Critical Phase of {{FeO}} Spans Conditions of {{Earth}}'s Lower Mantle},
  author = {Ho, Wai-Ga D. and Zhang, Peng and Haule, Kristjan and Jackson, Jennifer M. and Dobrosavljevi{\'c}, Vladimir and Dobrosavljevic, Vasilije V.},
  year = 2024,
  month = apr,
  journal = {Nature Communications},
  volume = {15},
  number = {1},
  pages = {3461},
  publisher = {Nature Publishing Group},
  issn = {2041-1723},
  doi = {10.1038/s41467-024-47489-w},
  urldate = {2026-02-18},
  copyright = {2024 The Author(s)},
  langid = {english}
}

@article{OhtaEtAl2012,
  title = {Experimental and {{Theoretical Evidence}} for {{Pressure-Induced Metallization}} in {{FeO}} with {{Rocksalt-Type Structure}}},
  author = {Ohta, Kenji and Cohen, R. E. and Hirose, Kei and Haule, Kristjan and Shimizu, Katsuya and Ohishi, Yasuo},
  year = 2012,
  month = jan,
  journal = {Physical Review Letters},
  volume = {108},
  number = {2},
  pages = {026403},
  issn = {0031-9007, 1079-7114},
  doi = {10.1103/PhysRevLett.108.026403},
  urldate = {2026-02-18},
  copyright = {http://link.aps.org/licenses/aps-default-license},
  langid = {english}
}

@article{OhtaEtAl2014,
  title = {Highly Conductive Iron-rich ({{Mg}},{{Fe}}){{O}} Magnesiow\"ustite and Its Stability in the {{Earth}}'s Lower Mantle},
  author = {Ohta, K. and Fujino, K. and Kuwayama, Y. and Kondo, T. and Shimizu, K. and Ohishi, Y.},
  year = 2014,
  month = jun,
  journal = {Journal of Geophysical Research: Solid Earth},
  volume = {119},
  number = {6},
  pages = {4656--4665},
  issn = {2169-9313, 2169-9356},
  doi = {10.1002/2014JB010972},
  urldate = {2026-02-18},
  copyright = {http://onlinelibrary.wiley.com/termsAndConditions\#vor},
  langid = {english}
}

@article{WuWahr1997,
  title = {Effects of Non-Hydrostatic Core-Mantle Boundary Topography and Core Dynamics on {{Earth}} Rotation},
  author = {Wu, X P and Wahr, J M},
  year = 1997,
  journal = {Geophysical Journal International},
  volume = {128},
  number = {1},
  pages = {18--42},
  issn = {0956-540X},
  isbn = {0956-540X}
}

@article{PuicaEtAl2023,
  title = {Analytical Computation of Total Topographic Torque at the Core--Mantle Boundary and Its Impact on Tidally Driven Length-of-Day Variations},
  author = {Puica, M and Dehant, V and Folgueira, M and Van~Hoolst, T and Rekier, J},
  year = 2023,
  month = jul,
  journal = {Geophysical Journal International},
  volume = {234},
  number = {1},
  pages = {585--596},
  issn = {0956-540X},
  doi = {10.1093/gji/ggad077},
  urldate = {2023-09-25},
  copyright = {All rights reserved}
}

@article{DehantEtAl2025,
  title = {Analytical Computation of the Total Topographic Torque at the Core--Mantle Boundary and Its Impact on Nutations},
  author = {Dehant, V and Puica, M and {Folgueira-L{\'o}pez}, M and Rekier, J and Van~Hoolst, T},
  year = 2025,
  month = apr,
  journal = {Geophysical Journal International},
  volume = {241},
  number = {1},
  pages = {474--494},
  issn = {1365-246X},
  doi = {10.1093/gji/ggaf050},
  urldate = {2026-02-20}
}

@misc{NIST:DLMF,
  key          = "{\relax DLMF}",
  year         = "2025",
  title        = "{\it NIST Digital Library of Mathematical Functions}",
  howpublished = "\url{https://dlmf.nist.gov/}, Release 1.2.5 of 2025-12-15",
  url          = "https://dlmf.nist.gov/5",
  note         = "F.~W.~J. Olver, A.~B. {Olde Daalhuis}, D.~W. Lozier,
                  B.~I. Schneider, R.~F. Boisvert, C.~W. Clark, B.~R. Miller,
                  B.~V. Saunders, H.~S. Cohl, and M.~A. McClain, eds."
}

@article{MandeaEtAl2015,
  title = {Gravimetric and Magnetic Anomalies Produced by Dissolution-Crystallization at the Core-Mantle Boundary},
  author = {Mandea, Mioara and Narteau, Cl{\'e}ment and Panet, Isabelle and Le Mou{\"e}l, Jean Louis},
  year = 2015,
  journal = {Journal of Geophysical Research: Solid Earth},
  volume = {120},
  number = {9},
  pages = {5983--6000},
  issn = {21699356},
  doi = {10.1002/2015JB012048}
}

@article{LeMouelEtAl2006,
  title = {Dissipation at the Core-Mantle Boundary on a Small-Scale Topography},
  author = {Le Mou{\"e}l, J. L. and Narteau, C. and {Greff-Lefftz}, M. and Holschneider, M.},
  year = 2006,
  journal = {Journal of Geophysical Research: Solid Earth},
  volume = {111},
  number = {B4},
  issn = {2156-2202},
  doi = {10.1029/2005JB003846},
  urldate = {2026-04-29},
  copyright = {Copyright 2006 by the American Geophysical Union.},
  langid = {english}
}

@article{SoldatiEtAl2012,
  title = {Tomography of Core--Mantle Boundary and Lowermost Mantle Coupled by Geodynamics},
  author = {Soldati, Gaia and Boschi, Lapo and Forte, Alessandro M.},
  year = 2012,
  month = may,
  journal = {Geophysical Journal International},
  volume = {189},
  number = {2},
  pages = {730--746},
  issn = {0956-540X},
  doi = {10.1111/j.1365-246X.2012.05413.x},
  urldate = {2026-02-27}
}

@article{BegganEtAl2026,
  author  = {Beggan, C. D. and others},
  title   = {International geomagnetic reference field: the fourteenth generation},
  journal = {Earth, Planets and Space},
  volume  = {78},
  pages   = {127},
  year    = {2026},
  doi     = {10.1186/s40623-025-02360-0}
}

@article{Aubert2026,
  title = {Geodynamo Simulations Spanning Millennia in the Physical Conditions of {{Earth}}'s Core},
  author = {Aubert, Julien},
  year = 2026,
  month = may,
  journal = {Journal of Studies of Earth's Deep Interior},
  volume = {2},
  publisher = {Episciences.org},
  issn = {3099-2877},
  doi = {10.46298/jsedi.17790},
  urldate = {2026-08-03},
  langid = {english}
}

@article{TylerKuang2014,
  title = {Resonant Tidal Excitation of Internal Waves in the {{Earth}}'s Fluid Core},
  author = {Tyler, Robert H. and Kuang, Weijia},
  year = 2014,
  month = jul,
  journal = {Physics of the Earth and Planetary Interiors},
  volume = {232},
  pages = {15--25},
  issn = {00319201},
  doi = {10.1016/j.pepi.2014.03.006},
  urldate = {2026-08-03},
  langid = {english}
}

@article{GilletEtAl2015,
  title = {Planetary Gyre, Time-Dependent Eddies, Torsional Waves, and Equatorial Jets at the {{Earth}}'s Core Surface},
  author = {Gillet, N. and Jault, D. and Finlay, C. C.},
  year = 2015,
  journal = {Journal of Geophysical Research: Solid Earth},
  volume = {120},
  number = {6},
  pages = {3991--4013},
  issn = {2169-9356},
  doi = {10.1002/2014JB011786},
  urldate = {2023-01-12},
  langid = {english}
}

@article{TrianaEtAl2021a,
  title = {The {{Viscous}} and {{Ohmic Damping}} of the {{Earth}}'s {{Free Core Nutation}}},
  author = {Triana, Santiago Andr{\'e}s and Trinh, Antony and Rekier, J{\'e}r{\'e}my and Zhu, Ping and Dehant, V{\'e}ronique},
  year = 2021,
  journal = {Journal of Geophysical Research: Solid Earth},
  volume = {126},
  number = {4},
  pages = {1--14},
  issn = {2169-9313},
  doi = {10.1029/2020jb021042},
  copyright = {All rights reserved}
}
\nocite{Aubert2026,BegganEtAl2026,TylerKuang2014,GilletEtAl2015,TrianaEtAl2021a}
%


%
%
%
%
%

\end{document}